\documentclass[useAMS,usenatbib]{mn2e}
\usepackage[pdftex]{graphicx}
\usepackage{amssymb,amsmath} 

\title[Near-infrared astrometry in the GC]{What is limiting near-infrared astrometry in the Galactic Center?}
\author[T. Fritz, S. Gillessen et al.]{T. Fritz$^{1}$, S. Gillessen$^{1}$, S. Trippe$^2$, T. Ott$^1$, H. Bartko$^1$, O. Pfuhl$^1$,\newauthor K. Dodds-Eden$^1$, R. Davies$^1$, F. Eisenhauer$^1$, and R. Genzel$^{1,3}$
\\
~\\
$^{1}$Max-Planck-Institute for Extraterrestrial Physics, 85748 Garching, Germany\\
$^{2}$IRAM, Grenoble, France\\
$^{3}$Department of Physics, University of California, Berkeley, 366 Le Comte Hall, Berkeley, CA94720-7300\\
}
\begin{document}

\date{Draft version Aug 2009}

\pagerange{\pageref{firstpage}--\pageref{lastpage}} \pubyear{2009}

\maketitle

\label{firstpage}

\begin{abstract}
We systematically investigate the error sources for high-precision astrometry from adaptive optics based near-infrared imaging data. We focus on the application in the crowded stellar field in the Galactic Center.
We show that at the level of $\lesssim 100\,\mu$as a number of effects are limiting the accuracy. Most important are the imperfectly subtracted seeing halos of neighboring stars, residual image distortions and unrecognized confusion of the target source with fainter sources in the background.
Further contributors to the error budget are the uncertainty in estimating the point spread function, the signal-to-noise ratio induced statistical uncertainty, coordinate transformation errors,   
the chromaticity of refraction in Earth's atmosphere, the post adaptive optics differential tilt jitter and anisoplanatism. For stars as bright as $m_\mathrm{K}=14$, residual image distortions limit the astrometry, for fainter stars the limitation is set by the seeing halos of the surrounding stars. In order to improve the astrometry substantially at the current generation of telescopes, an adaptive optics system with high performance and weak seeing halos over a relatively small field ($r\lesssim3''$) is suited best. Furthermore, techniques to estimate or reconstruct the seeing halo could be promising.
\end{abstract}


\section{Introduction}

The Galactic Center (GC) is a unique celestial laboratory. It hosts a massive black hole (MBH) the existence of which is now widely accepted in astronomy \citep{Wollman:1977p1438,Genzel:1996p199,Ghez:1998p118,Schodel:2002p153}. 
The tightest constraints on the mass have been put by observing the motions of individual stars as they orbit the MBH \citep{Ghez:2008p945,Gillessen:2009p1117}. The statistical error of the mass is as low as $\approx\,$1.5\%, the systematic error is considerably higher with $\approx\,$10\% owed to the uncertainty in $R_0$ of $\approx\,$5\%, the distance to the GC.

These astonishing measurements rely on high-angular resolution, adaptive optics (AO) assisted, near-infrared observations, obtained at large telescopes; namely the Keck and VLT facilities. From imaging data astrometric positions are derived, reaching an accuracy of $\approx 300\,\mu$as. This a factor of 200 smaller than the maximum possible image resolution of $\approx60\,$mas in K-band, while the apparent orbit of the most important star of these so-called S-stars, S2, measures $\approx190\,$mas. 

With the black hole paradigm now being well established, the efforts turn towards detecting post-Newtonian effects. From General Relativity, one expects a prograde periastron shift of $0.2^\circ$ per revolution of 15.8 years for S2 \citep{Rubilar:2001p186}. This corresponds to an apparent position shift of $\approx800\,\mu$as, which however is not as easily detected as it seems, since it needs to be measured from the same data from which also the orbital elements have to be determined. Secondly, a population of dark stellar remnants such as stellar mass black holes and neutron stars might be present around the MBH \citep{Morris:1993p127}. Such an extended mass component will lead to a retrograde periastron shift. If for example $\approx 0.1$\% of the mass inside the
S2 orbit were extended, the resulting retrograde shift would cancel the relativistic prograde precession \citep{Rubilar:2001p186}. Detecting either of these shifts would be astrophysically extremely interesting.

Future ground-based telescopes will further increase the spatial resolution accessible, thus potentially increasing the astrometric accuracy further. Trippe et al. (in prep.) investigate in general the astrometric performance of an extremely large telescope.  \cite{Weinberg:2005p140} simulated the expected advance in monitoring stellar orbits in the GC when using an extremely large telescope, concluding that the upcoming facilities indeed will be capable of measuring the post-Newtonian effects and that mass of and distance to Sgr~A* can be determined with unprecedented 
precision.  

High-precision astrometry is important for other scientific cases, too. Intermediate mass black holes might be found by detecting accelerations of stars in the vicinity of the black hole. Potential sites are for example $\omega$ Cen \citep{Noyola:2008p2001,Anderson:2009p1233} or the compact stellar group GC IRS13E (\cite{Maillard:2004p121,Schodel:2005p185}; Fritz et al. (in prep.)). Another area where AO-assisted high-resolution imaging profits from excellent astrometric capabilities is the domain of binaries and substellar companions (\cite{Chauvin:2004p2179,Neuhauser:2007p2004, Kohler:2008p2183,Lagrange:2008p2177,Neuhauser:2008p2002}; Kervella et al. (in prep.)). 

This paper aims at investigating the limits of astrometry, in particular in the GC. This is of great value both for current and future studies of stellar orbits. Our approach is to estimate the various error sources mostly empirically, i.e. from existing VLT data. The main parameters characterizing the reachable accuracy for any given star are its magnitude and distance from Sgr~A*, and for any given image the Strehl ratio. The empirical numbers used were determined for the VLT NIR-AO imager NACO, operated in K$_S$-band with a pixel scale of $13\,$mas/pix. The field of view being of interest here is $\Theta_\mathrm{FoI} = \pm 2''$, given by the apparent size of the GC stellar system. We concentrate on the data set from 13 March 2008, which was of good quality (Strehl ratio larger than 30\%) and therefore is well-suited to study systematic effects.\footnote{We are using data from ESO program 179.B-0261.} In that sense, this paper is a case study, but the dependencies on the Strehl ratio which we exploit make the results applicable for a wider range of data.

\section{Statistical Uncertainty}
\label{s2}

\begin{figure}
\includegraphics[width=80mm]{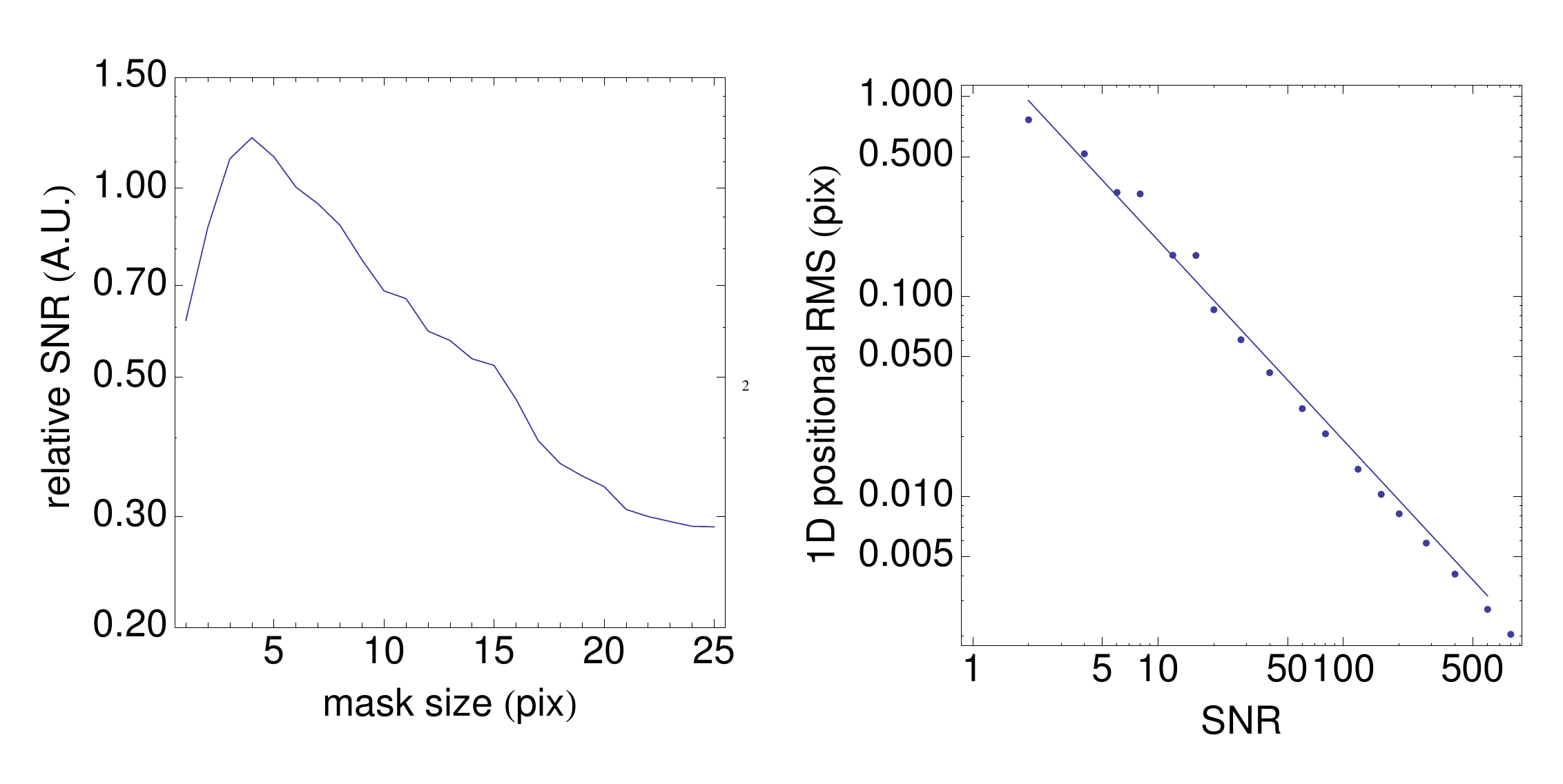}
  \caption{Left: SNR for a source with FWHM$\,=6\,$pix and Gaussian shape in a noise image as a function of the integrating radius used, showing that a source area with $r=4\,$pix is optimal. Right: The data points show the positional uncertainty in simulated images as function of the SNR. The solid line is the prediction from equation~\ref{e1}.}
\label{f1}
\end{figure}

The fundamental limit to astrometry for diffraction limited data is given by the number of photons recorded and the image resolution, with higher SNR and smaller point spread function (PSF, FWHM $\Theta$) increasing the positional accuracy. For a circular aperture of size $D$ and an observation wavelength $\lambda$ one has
\begin{equation}
\Theta_\mathrm{FWHM}= 1.028 \frac{\lambda}{D}\,\,.
\label{e1}
\end{equation}
\cite{Lindegren:1978p1953} gives the following formula for the position error:
\begin{equation}
\sigma_x = \frac{1}{\pi}\, \frac{\lambda}{D}\,\frac{1}{\mathrm{SNR}}\,\,.
\label{e1}
\end{equation}
We have verified this formula explicitly by simulating different SNR for a simple Gaussian PSF. The sampling was chosen to be similar to VLT data from imaging GC observations with $\Theta_\mathrm{FWHM} \approx 6\,$pix which occurs for a Strehl ratio of $18\%$\footnote{This is taken from the actual data, spanning the years 2002-2009. Since 2007 we are getting consistently smaller FWHM values and higher Strehl ratios.}. We generated a set of noise images, in each of which the PSF was added. First, we searched the optimum radius for the source region for the given sampling. We determined the SNR as ratio of flux inside a certain radius and the standard deviation of the flux inside the same area in the noise image. The optimum was 4 pixels (figure~\ref{f1}, left), independent from the peak intensity. Next, we varied the SNR by adding the PSF 
several times to each noise map with the peak intensity ranging from $0.3\,\times$ to $200\,\times$ the noise level. We determined the position of the PSF by a Gaussian fit to each image. The standard deviation over the set of noise images per intensity then measures the positional uncertainty for that intensity level. The corresponding SNR was calculated with the optimum radius. We plot the positional error as a function of SNR in figure~\ref{f1}, right, which shows that our explicit simulation matches equation~\ref{e1}. 
While the 1/SNR behavior is obvious, this checks mainly that the numerical factor in equation~\ref{e1} correctly describes the accuracy for the optimum signal extraction.

Empirically, we find for the VLT data that a source in K-band creates a total current (photo electrons per second) on the detector (in its high dynamic mode with the readout scheme double read-reset-read) of
\begin{equation}
\Sigma_\mathrm{pix} \dot{n}_\mathrm{e,\,obj}[\mathrm{el/sec}] = 10^{0.4\, (25.4 - m_\mathrm{K})}\,\,
\label{e2}
\end{equation}
If source photon noise were the only error source, for a $m_\mathrm{K}=14$ source (i.e. S2), within one second the position uncertainty would drop to $50\,\mu$as. However, also the noise of the sky background and the read noise of the detector have to be taken into account in equation~\ref{e1}.

In a single pixel in a single-pointing, combined frame the noise after the usual data reduction steps of sky subtraction and flat fielding should be (assuming that the flat field on average equals 1 and that $N_\mathrm{sky}$ frames with the same exposure time as the $N_\mathrm{obj}$ object frames were averaged for the sky)
\begin{equation}
\sigma_\mathrm{e}^2 = {\sigma_\mathrm{f}^2 s^2} + \frac{n_\mathrm{e,\, obj}}{N_\mathrm{obj}}  +\left(\frac{1}{N_\mathrm{obj}}+\frac{1}{N_\mathrm{sky}}\right)\,\times \, (\sigma_\mathrm{RON}^2+n_\mathrm{e,\, sky}),
\label{e3}
\end{equation}
where $\sigma_\mathrm{f} = 0.15\%$ is the empirically determined relative noise of the flat field, $s = (n_\mathrm{e,\, obj} -  n_\mathrm{e,\, sky})/f$ the reduced signal per pixel and $\sigma_\mathrm{RON}=46.2$ the read noise of the detector in electrons. Typically, the sky brightness per pixel for the given setup is
\begin{equation}
\dot{n}_\mathrm{e,\,sky}[\mathrm{el/sec}] = 21.5\,\,.
\label{e4}
\end{equation}

\begin{figure}
\includegraphics[width=80mm]{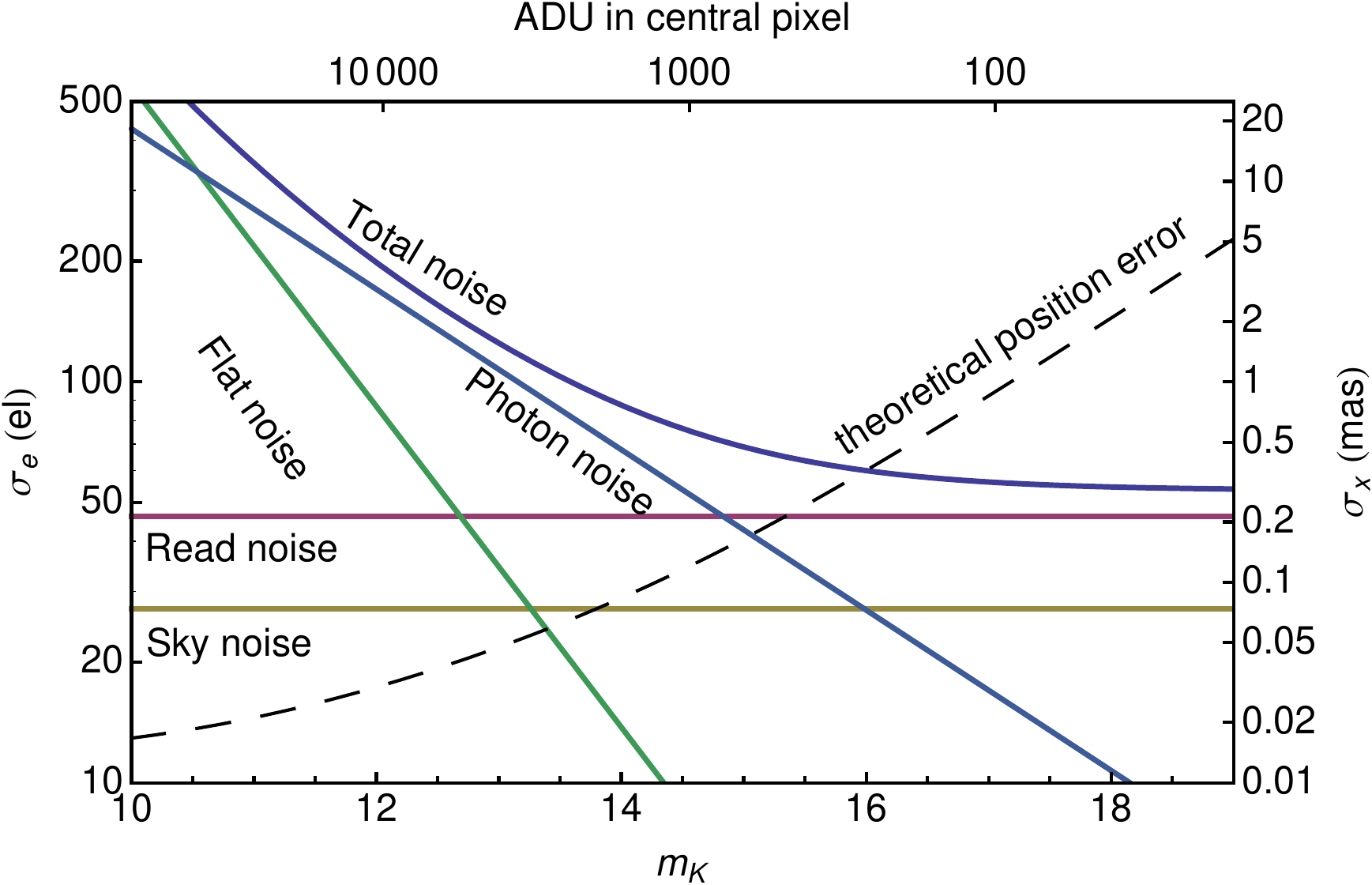}
  \caption{The solid lines (referring to the left ordinate) show the noise contributions per pixel as function of illumination according to equation~\ref{e3} for a single image, which actually is the combination of two object and two sky frames of 17.2 sec each. (Our data is mostly obtained in a mode with 2 detector reads of 17.2 sec per data file, a setting that is chosen to avoid saturation effects for most of the stars and minimizes overheads.) The total noise at faint magnitudes is dominated by the read noise. For realistic data sets, we use more than two sky frames, such that the read noise contribution from the sky is reduced compared to this illustration. For brighter objects, the photon noise is the dominant contribution. The regime in which the noise of the flat becomes important already is in the highly non-linear regime of the detector which normally is not used. For the conversion to magnitudes, a sampling of $6\,$pix per FWHM and a Gaussian PSF that contains $30\%$ of the light were assumed. The dashed line (referring to the right ordinate) gives the resulting positional error according to equation~\ref{e1} for the total noise.}
\label{f2}
\end{figure}

The solid lines in figure~\ref{f2} illustrate the noise behavior for a single pixel as given by equation~\ref{e3}. With the noise (equation~\ref{e3}) and the signal (equation~\ref{e2}), it is straight forward to evaluate equation~\ref{e1} if one assumes a certain PSF shape. As a simple example, we show the resulting positional uncertainty for a case in which the position is estimated from a Gaussian shaped PSF-core with a FWHM of 6 pixels that contains $30\%$ of the stellar light. As above, the light from inside a 4-pixel radius was used. The error so obtained represents the statistical limit to the positional accuracy for a single, reduced frame. Usually, astrometry is done on combined objects frames with varying pointing positions ('mosaics'). This improves the statistical precision limit by a factor $\sqrt{N_\mathrm{pointing}}$ compared to equation~\ref{e3}.

\section{Astrometry in the GC}
We briefly describe the method by which we obtain astrometric positions. The lack of any extragalactic background source in the NIR is one of the main obstacles for astrometry in the GC. Therefore all position measurements are relative to other sources in any given image. The link to the international celestial reference frame ICRF is only possible due to a set of SiO maser stars, which are both NIR and radio sources. The position vectors relative to Sgr~A* of the latter can be measured with high accuracy \citep{Reid:2003p142, Reid:2007p169}. The position and motion of Sgr~A* in the ICRF in turn is well known \citep{Reid:1999p1531,Reid:2004p190}. 

An additional complication is that with the current 1k$\times$1k NACO detector, a suitable sampling ($\approx4\,$pix$/$FWHM$\,=13\,$mas/pix, Trippe et al. in preparation) is only reached for the central few arcseconds (depending on the dithering scheme), while the SiO maser stars are found out to 20''. Therefore it is more practical to measure the SiO maser positions with a larger pixel scale ($27\,$mas/pix for NACO) and relate the finer scale to the coarser sampling by a set of reference stars that can be reliably detected in both scales. That also helps to overcome the large dynamic range needed, given that the brightest SiO maser star used for astrometry has a magnitude of $m_\mathrm{K}\approx8.5$ \citep{Blum:2003p1118, Reid:2007p169} and that one is interested in the positions of stars as faint as $m_\mathrm{K}\approx19$.

In practical terms, the procedure is as follows: From a set of images, obtained between 2002 and 2009 in the $27\,$mas/pix scale, we derive astrometric positions and proper motions for a set of $\approx 100$ reference stars. This relies on the work of  \cite{Reid:2007p169}, which allows us to calculate the astrometric positions of the SiO maser stars for the given NIR epochs. We use mosaics corrected for their geometric distortion \citep{Trippe:2008p1123} and a full, 6-parameter linear transformation to relate pixel and astrometric positions of eight SiO masers. For any given image in the $13\,$mas/pix scale we determine the PSF from the image and deconvolve it using the Lucy-Richardson algorithm \citep{Lucy:1974p216}. After beam restoration with a Gaussian beam we determine the stellar pixel positions by Gaussian fits to the positions, both to the reference stars and the sources targeted. The transformation used to link the astrometric reference star positions to their pixel positions is a 20-parameter, third order polynomial transformation, which should also implicitly correct for any large scale (5'') image distortion. This is useful since for the smaller pixel scale we were not able to construct a reliable distortion model, indicative that the effect is fairly small \citep{Trippe:2008p1123}. For a more complete description of the procedure see \cite{Gillessen:2009p1117}.

\section{Systematic Uncertainties}
A multitude of systematic uncertainties are present in the astrometric data, the most important ones being atmospheric turbulence, image distortions, unrecognized source confusion and uncertain PSF halos.
The time scales involved are very different; atmospheric effects are present even in a single frame, while confusion of sources happens on time scales of years. The section discusses the error sources (roughly) by increasing time scale involved.

\subsection{Fast atmospheric Limitations}

\subsubsection{Anisoplanatism}
Time variable refraction in Earth's atmosphere, induced by turbulence cells, blurs ground-based astronomical images, an effect which is called 'seeing'. For the observations discussed here, the seeing is partly corrected by the adaptive optics (AO) system. The resulting PSF is a superposition of a close to diffraction-limited core with a seeing-limited halo. Useful parameters to describe the performance of the AO are the Strehl ratio (SR; the ratio of measured central flux compared to the diffraction-limited central flux) and the FWHM of the PSF. The AO correction in our data is achieved with a single guide star (single-conjugate AO). 

The tip-tilt component of the wavefront errors between any two objects in the field of view gets less and less correlated with increasing distance between the two objects, an effect which is called anisoplanatism. Complete decorrelation is reached for the so-called isoplanatic angle, which has a typical value of 
$10'' - 20''$. As a result of anisoplanatims, one observes random variations in the relative positions of any two objects. 
Anisoplanatism is seen as PSF elongation during individual exposures, and as residual, differential tilt jitter between successive frames.
  
\subsubsection{Differential Tilt Jitter}
The differential tilt jitter between any two stars in good approximation linearly increases with distance between the two stars. The effect averages out with exposure time.
For a Komolgorov turbulence spectrum the dominant term of the differential tilt jitter scales with telescope diameter $D$ and integration time $t$ like \citep{Cameron:2009p1964}
\begin{equation}
\left(
\begin{array}{c}
 \sigma^2_ {\parallel,\, \mathrm{TJ}}\vspace{0.1cm}\\
 \sigma^2_{\perp,\,\mathrm{TJ}}   
\end{array} 
\right) =\,\alpha 
\left(
\begin{array}{c}
 3\vspace{0.1cm}\\
1 \end{array} \right)\,
\theta^2 D^{-7/3} \,\frac{\tau}{t}\,
\,.
\label{e5}
\end{equation}
The effect is $\sqrt{3}\times$ bigger in the direction connecting the two stars than perpendicular to it. The time constant $\tau$ characterizes the time it takes a turbulence cell to move over the telescope aperture. The constant $\alpha$ is related to the second moment of the atmospheric turbulence profile. A typical value, following from the numbers in \cite{Cameron:2009p1964}, is $\alpha\approx 3$ when $\sigma$ is measured in mas, $D$ in meters and $\theta$ in arcseconds.

\begin{figure}
\includegraphics[width=80mm]{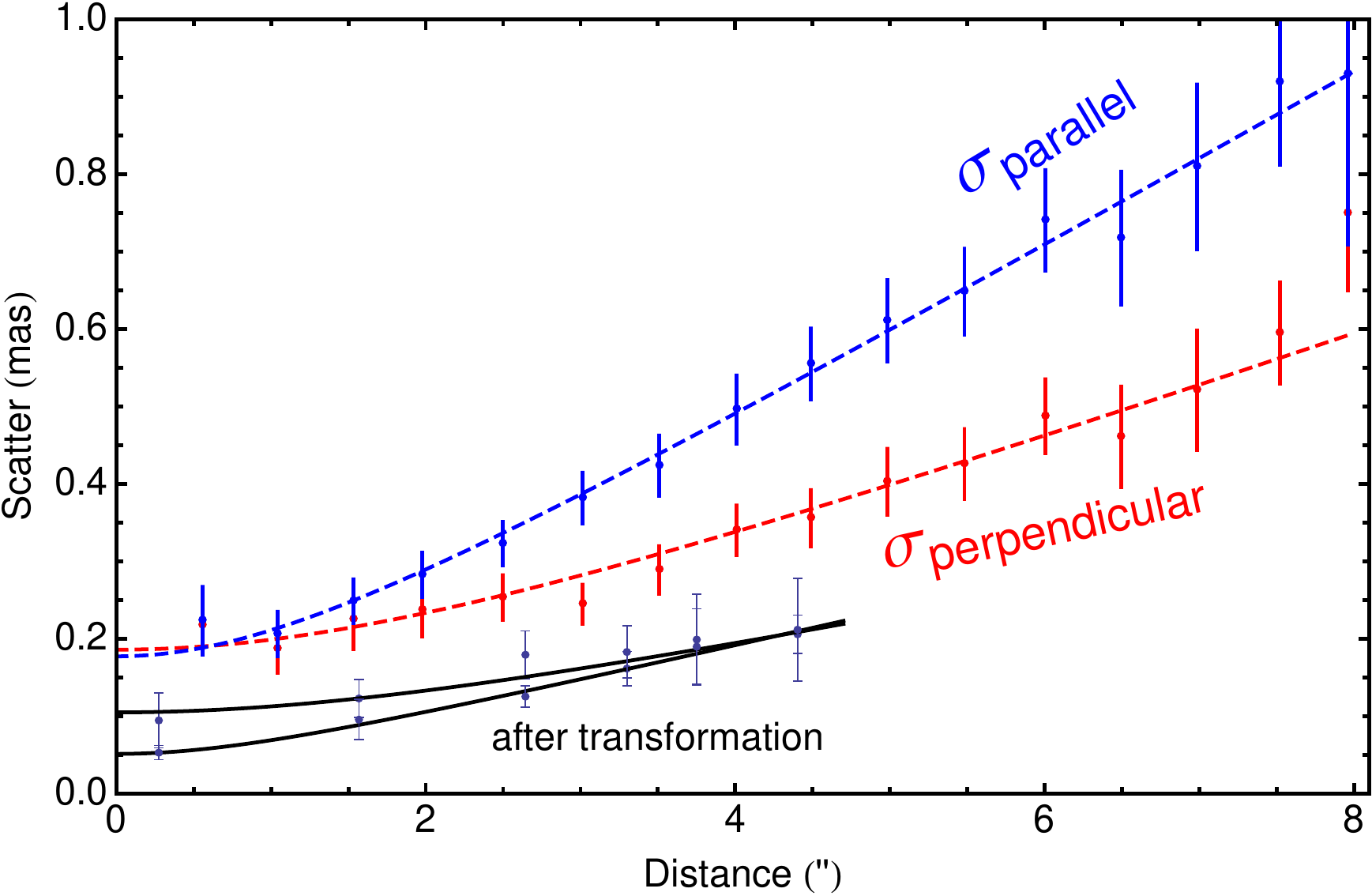}
  \caption{Differential tilt jitter in the data from 13 March 2008. The dashed lines show linear fits to the positional scatter between pairs of stars in the direction parallel or perpendicular to difference vector between the two stars. The empirical model used was
 $\sigma^2 = \sigma_0^2 + \beta^2 \theta^2$. The fitted slopes $\beta$ for large distances are $0.114\pm0.002\,\mathrm{mas}/''$ (blue) and $0.071\pm0.003\, \mathrm{mas}/''$ (red). The solid, black line and corresponding data show the remaining astrometric effect in R.A. and Dec. after a full linear transformation (the range is smaller since positions are counted from the center of the field then). }
\label{f3}
\end{figure}

We looked for the effect in the data set from 13 March 2008 by calculating the variance of stellar distances in a sequence of subsequent exposures that were obtained at exactly the same pointing position. We decomposed the difference vectors into the directions parallel and perpendicular to the vector connecting the two respective stars. Figure~\ref{f3} (dashed lines) shows the resulting scatter as a function of distance between the two stars. Clearly the differential tilt jitter is visible and the measured ratio $\sigma_ \parallel/\sigma_ \perp=1.62 \pm 0.05$ is consistent with the expected value of $\sqrt{3}$. Since we do not have the data to evaluate the turbulence profile for our observation, we can check for plausibility only. From the slope of the relation $d\sigma_\mathrm{TJ}/d\theta=0.071\pm0.03$ and given the exposure time of $34.4\,$s we estimate the wind crossing time over the telescope aperture of $8\,$m to be $0.4\,$s, corresponding to a wind speed of $v\approx20\,$m/s, which seems reasonable.

Since the field of interest $\Theta_\mathrm{FoI} = \pm 2''$ is practically always smaller than the isoplanatic angle,
the tilt jitter will act in a correlated way on all of the stars. Hence, the effect will to a large extent cancel out for astrometric positions defined as relative positions to a set of reference stars \citep{Cameron:2009p1964}. The effect of the corresponding full linear transformation (i.e. including shear terms) is shown for the test data in figure~\ref{f3} (solid lines).
The remaining astrometric error for a single frame is $< 0.2\,$mas and in the center even $< 0.1\,$mas. Due to the mosaicking, that number will decrease further by $\sqrt{N_\mathrm{frames}}$ and for a typical data set of $\gtrsim70$ images differential tilt jitter should contribute less than $20\,\mu$as. Hence, using a simple linear transformation is sufficient to effectively eliminate this error source. The more elaborate approach by \cite{Cameron:2009p1964} is not necessary for our current GC data sets.

\subsubsection{PSF elongation}
The AO corrects the tip-tilt jitter optimally for the the guide star. Since the exposure times are much longer than the time constant for the AO, the image retrieved can be thought as a superposition of many short images, each of which is centered on the guide star. As a result, the random errors from equation~\ref{e5} vanish at the guide star and the image is sharpest there. Further away, the integrated jitter increases the PSF width with distance to the guide star, and the particular form of equation~\ref{e5} leads to an elongation of the PSF in direction towards the guide star. Since the analysis uses a constant PSF (either for deconvolution or PSF fitting) when estimating stellar positions, the effect will lead to increased position errors.

We assessed the error by comparing the differences of extracted position from various PSF estimates, obtained from stellar images with similar brightnesses at different positions in the field of view. We used the high-quality data set from 13 March 2008. Using two different PSF estimates from the same region leads to a typical position difference of $50\,\mu$as. When the two PSF estimates are extracted from two samples $3.5''$ apart, we only find a small increase of the typical position difference to $70\,\mu$as. At $6''$ distance, the effect starts to hurt more severely since it reaches 
$\approx180\,\mu$as. At $7''$ distance, it even reaches $400\,\mu$as. These values illustrate the effect of the isoplanatic angle. It is worth noting that anisoplanatism can be significantly more severe for worse atmospheric conditions.

For the analysis, the effect will contribute only at a very low level, if the set of stars from which the PSF is constructed is chosen carefully; i.e. within $\Theta_\mathrm{FoI}\approx\pm2''$ of the target region. In GC data sets, this can easily be achieved and therefore we expect anisoplanatism to be only a very small error contribution. If the size of the target region exceeds 4'', as for example in the work of
\cite{Trippe:2008p1123} or \cite{Schodel:2009p1100}, the anisoplanatic effects should be taken into account for astrometric measurements by varying the PSF over the field. 

\subsubsection{Comparison with data}

\begin{figure}
\includegraphics[width=80mm]{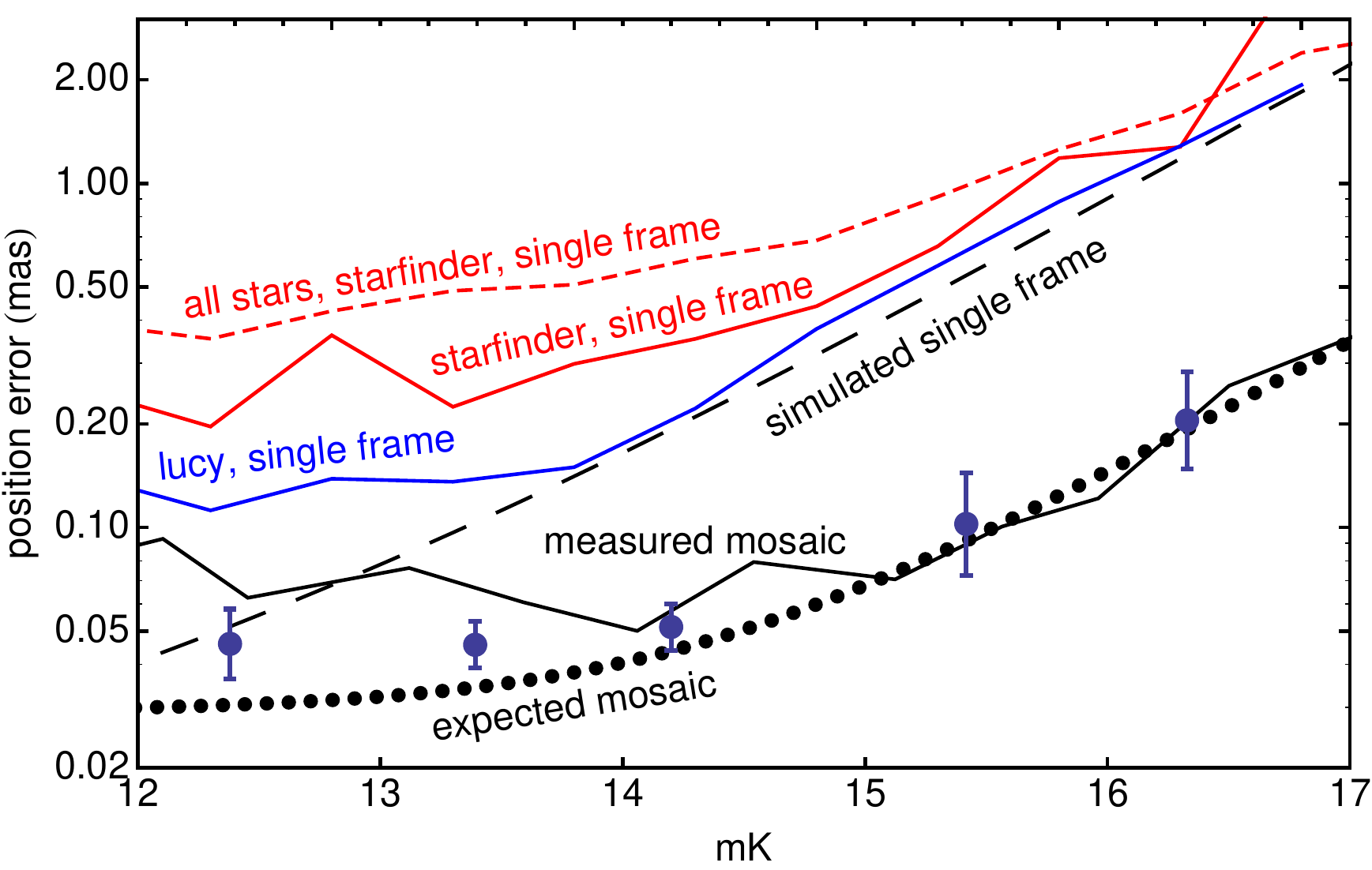}
  \caption{Empirical position errors as a function of stellar magnitude, estimated using a PSF extracted from the respective data. The red, dashed line indicates the errors for all stars found by starfinder in a single frame and is dominated by anisoplanatism at brighter magnitudes. The red, solid line shows the errors for isolated stars obtained from starfinder on the same single image in a restricted area (the central 2'', $\Theta_\mathrm{FoI}$). The blue line indicates for the same stars the errors obtained by applying a Lucy-Richardson deconvolution to the single frame, and fitting the sources with Gaussian functions. The black, dashed line shows the errors obtained by simulating isolated stars and finding them back. The black dotted line is showing the expectation for a mosaic, obtained by describing the blue line with an empirical function and square root scaling with the number of frames used in the mosaic. The black, solid line shows the measured errors on the mosaic which flatten at a level of $\approx 70\,\mu$as. The data points show the measured, statistical errors for the mosaic from 13 March 2008.}
\label{f4}
\end{figure}

Figure~\ref{f4} compares the expected position errors with the actual scatter in the data, using again the data set from 13 March 2008. We obtained the expected errors by simulating isolated stars and remeasuring their positions, using a noise floor as measured in single frames and using the respective PSF also obtained from the data. The position scatter between subsequent single frames is a useful estimator for the position error due to the statistical limit and atmospheric turbulence, since the frames were taken with exactly the same pointing position. This means neither image distortions nor confusion events bias the scatter.
 
We find that compared to the simulations, positions found by starfinder \citep{Diolaiti:2000p1984} scatter as expected from anisoplanatism. The scatter gets reduced, if one restricts the analysis to stars that are sufficiently isolated in a small area ($\Theta_\mathrm{FoI}=\pm 2''$). For $m_\mathrm{K}>15$, the scatter is only moderately higher than in the simulation. In comparison, our technique of Lucy-deconvolving the image performs a bit better than starfinder and is for $m_\mathrm{K}>14$ essentially as good as predicted by the simulation. The reason why starfinder is performing a bit worse might be that it was allowed to calculate a (spatially varying) background map for each single image, which however is not needed for the data. The additional degree of freedom then introduces extra noise, from which the deconvolution technique does not suffer. Both for starfinder and deconvolution, the brighter stars have a higher scatter than what one would expect from the simulations and the values reach a floor. Its level of $\approx125\,\mu$as is consistent with being the combination of PSF uncertainty (section~\ref{secPSF}) and differential tilt jitter as present in figure~\ref{f3} (data after transformation).  

The empirical curve for the deconvolution of a single frame was then parametrized by a simple model of type $\sigma_x^2 = \alpha\, \sigma^2(m_\mathrm{K})+\sigma_0^2$. For the given data, $\alpha=1.1$ and $\sigma_0=100\,\mu$as are an excellent description of the measured errors.
We scaled this relation by $\sqrt{N_\mathrm{frames}}=\sqrt{24}$ in order to estimate the expected positional uncertainty for a single-pointing mosaic (figure~\ref{f4}). For $m_\mathrm{K}>15$ the actual data (obtained again with deconvolution) follows the expected relation and flattens for sources with $m_\mathrm{K} \lesssim14$ at about $70\,\mu$as. This is the value as expected from the PSF uncertainty (section~\ref{secPSF}) and the anisoplanatism over the field. Hence, we conclude that no additional terms come in by combining multiple frames to a single-pointing mosaic.

For comparison, we also show in figure~\ref{f4} (data points) the positional errors as obtained by comparing two mosaics created from subsamples using the data from 13 March 2008. The match with the simulated is actually a combination of the fact that $1.5\times$ more single frames were used here for each submosaic, and that the error is composed of the fitting error plus the positional difference between the two submosaics.
 
\subsection{Slow atmospheric Limitations}

\subsubsection{Differential refraction}
Refraction in Earth's atmosphere displaces stellar images towards the zenith by
\begin{equation}
R\approx \frac{n_0^2-1}{2 n_0^2} \frac{p\,T_0}{p_0\,T} \tan z \approx 44'' \tan z\,\,,
\end{equation}
where $z$ is the zenith angle of the observation and the constant of $44''$ is the refraction constant for typical conditions at Paranal as evaluated from the refractive index $n_0$ at standard conditions of $p_0=1013.25\,$mbar and $T_0=273.15\,$K, scaled by the current pressure and temperature.
Since we are using relative astrometry, only the differential effect is relevant. It amounts to 
\begin{equation}
\delta R \approx \Theta_\mathrm{FoI} [\mathrm{rad}] \times 44''/\cos^2 z \,\,.
\end{equation}
Numerically, this evaluates to values around $3.5\,$mas for $z=60^\circ$ and $ \Theta_\mathrm{FoI}=\pm 2''$. It would thus be a large error source. However, the effect is absorbed into the shear terms of the transformation; a full linear transformation leaves a remaining second order effect that is less than $3\,\mu$as.

\subsubsection{Chromatic effects}
The dependence of the refractive index $n$ on the observation wavelength (e.g. as given by \cite{Edlen:1953p1997}) for the H- and K-band and typical Paranal conditions is approximated by
\begin{equation}
n_1\cdot\,10^7 =2029.94 - 2.87 \lambda_2 + 2.16 \lambda_2^2 - 
 1.44 \lambda_2^3 + 0.92  \lambda_2^4 \,\,
\end{equation}
where $n_1=n-1$ and $\lambda_2=\lambda[\mu\mathrm{m}]-2$. Thus the atmosphere acts as weak spectrometer, as a result of which the effective position for a source is the weighted average of the wavelength dependent positions over the band \citep{Helminiak:2009p2198}. The weighting factor is the number of photons per wavelength which in turn is the convolution of the input spectrum with the interstellar extinction and the atmospheric and instrumental transmission numbers. The size of the net effect grows quadratically with the bandpass used and hence narrow band filters help to suppress the atmospheric chromatic effects on the astrometry.

\begin{figure}
\includegraphics[width=80mm]{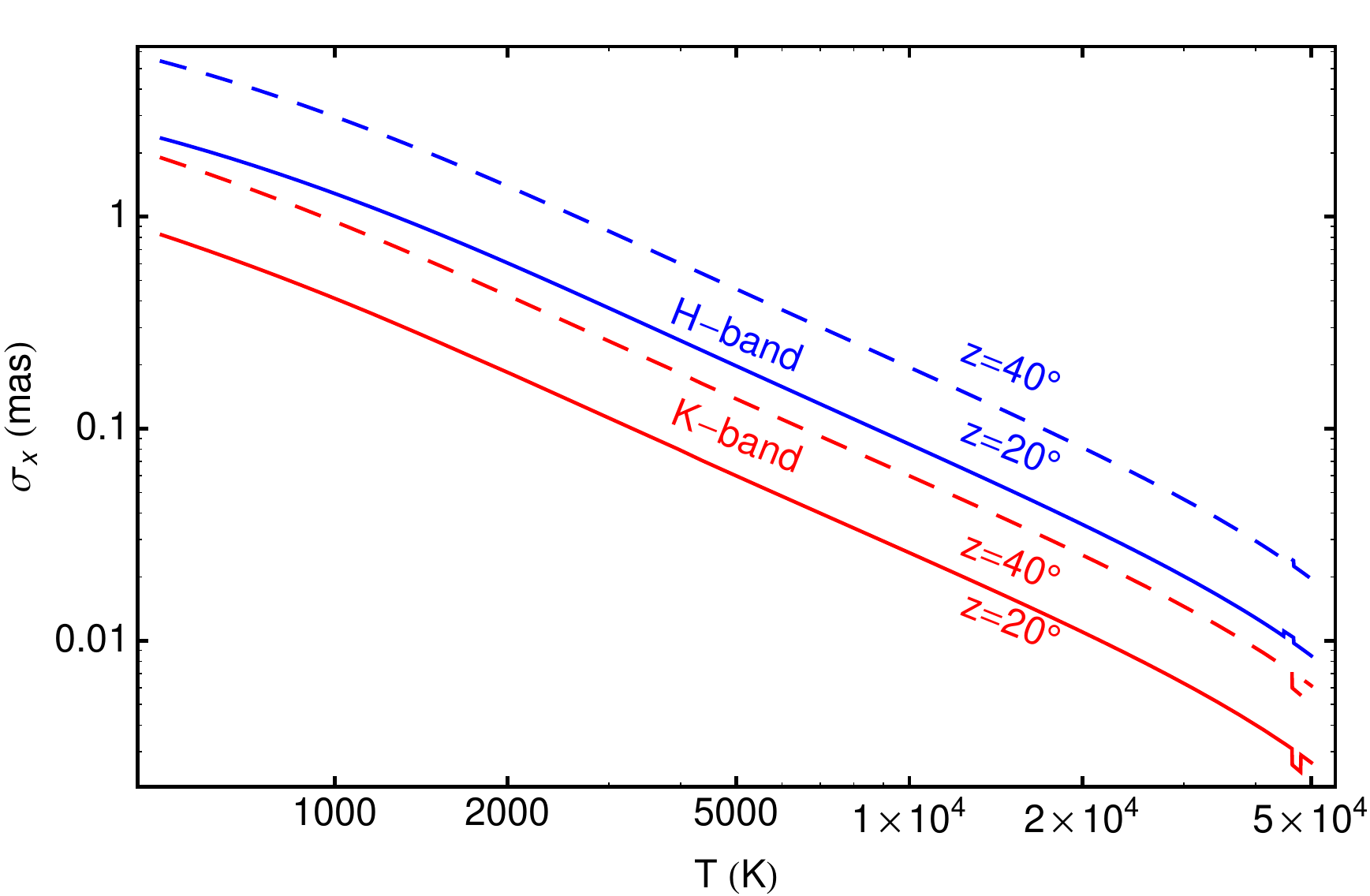}
  \caption{Assuming black body emission, the astrometric shift of a source due to the chromatic effects of Earth's atmosphere compared to a fictitious emitter of $T=10^5\,$K is shown as a function of temperature. The red lines assume a K-band measurement, the blue lines H-band. The solid lines are for a zenith angle of $20^\circ$, the dashed lines for $z=40^\circ$.}
\label{f5}
\end{figure}

Figure~\ref{f5} shows the resulting effect for black body type emission compared to a fictitious object with $10^5\,$K. For young, hot stars (most of the S-stars have $T\approx 25000\,$K, \cite{Ghez:2003p178,Martins:2008p177}) the K-band measured positions for typical zenith angles are altered by $\approx20\,\mu$as. For stars with a temperature of $6000\,$K the K-band shift would be $\approx70\,\mu$as. Below $5000\,$K, the approximation as black body breaks down due to the presence of the broad CO absorption features around $2.3\,\mu$m, and the shift does not get larger anymore but rather is close to 0 with a typical spread of $<\pm20\,\mu$as compared to the fictitious emitter. 

Since our position measurements are relative to other sources in the field, the differential effect between reference stars and target sources is an error source for the positions obtained. The reference frame is constructed from early- and late-type stars in roughly equal numbers, such that in K-band it refers to coordinates shifted by $\approx10\,\mu$as compared to the fictitious object. Individual stars are then shifted again depending on their stellar type by $\approx10\,\mu$as compared to the coordinate system. Only G-type stars would experience a shift of $\approx50\,\mu$as; but main sequence stars of that type are by far too faint to be detected in the GC field.
The net effect for K-band based astrometry is thus very small. 

For H-band data the effect is larger in the first place (figure~\ref{f5}) and secondly late-type stars  actually can be approximated by black bodies in H-band. For $z=40^\circ$ and a late-type star the chromatic shift can reach $400\,\mu$as compared to the $10^5\,$K emitter. For hot, young stars a more typical value is $50\,\mu$as. The astrometric net effect will thus be $\approx 175\,\mu$as.

Of course, these values strongly depend on the actual zenith angle and the exact input spectrum. Also, this error source is correctable, given that the zenith angle is known and if the input spectrum is at least approximately known.
  
\begin{figure}
\includegraphics[width=80mm]{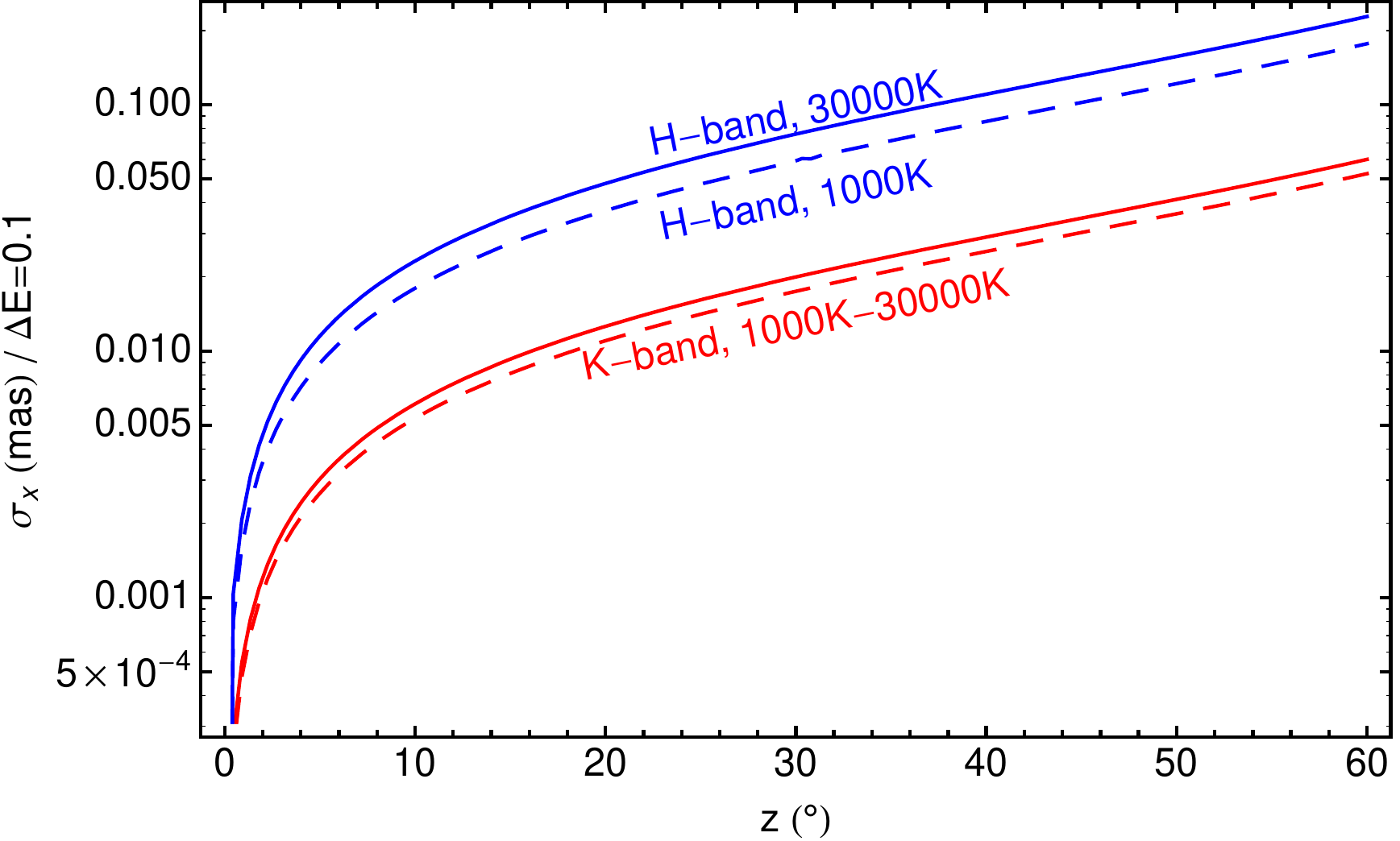}
  \caption{Positional shift of a black body source per $0.1\,$mag change in extinction as a function of zenith angle as it results from the chromaticity of Earth's atmosphere. The red lines show that the effect for K-band data is nearly independent from the source temperature. For H-band data (blue lines), the effect slightly increases with temperature.}
\label{f6}
\end{figure}

More subtle is the combined effect of the patchy nature of the extinction screen towards the GC
and the chromaticity of the atmosphere. Extinction variations lead to a change in observed color for each object as it moves behind the screen, the change in color in turn leads to a positional offset. By looking at the scatter of the observed K-band magnitudes of some isolated, bright stars (e.g. S8, S10, S30, S65, S87 in the nomenclature of \cite{Gillessen:2009p1117}) we conclude that the extinction variation $\Delta A_\mathrm{K} \lesssim 0.1$ for the central field of interest $\Theta_\mathrm{FoI}$. Figure~\ref{f6} plots the positional shift per 0.1 mag extinction variation as a function of zenith angle. For K-band data, the effect is $\sigma_x\lesssim 20\,\mu$as, for H-band data $\sigma_x\lesssim 100\,\mu$as. Also note that for larger regions in the GC of a few arcsecond, values of $\Delta E \approx 0.5$ are reported \citep{Buchholz:2009p1131}, which yields a correspondingly larger chromatic effect. 

\begin{figure}
\includegraphics[width=80mm]{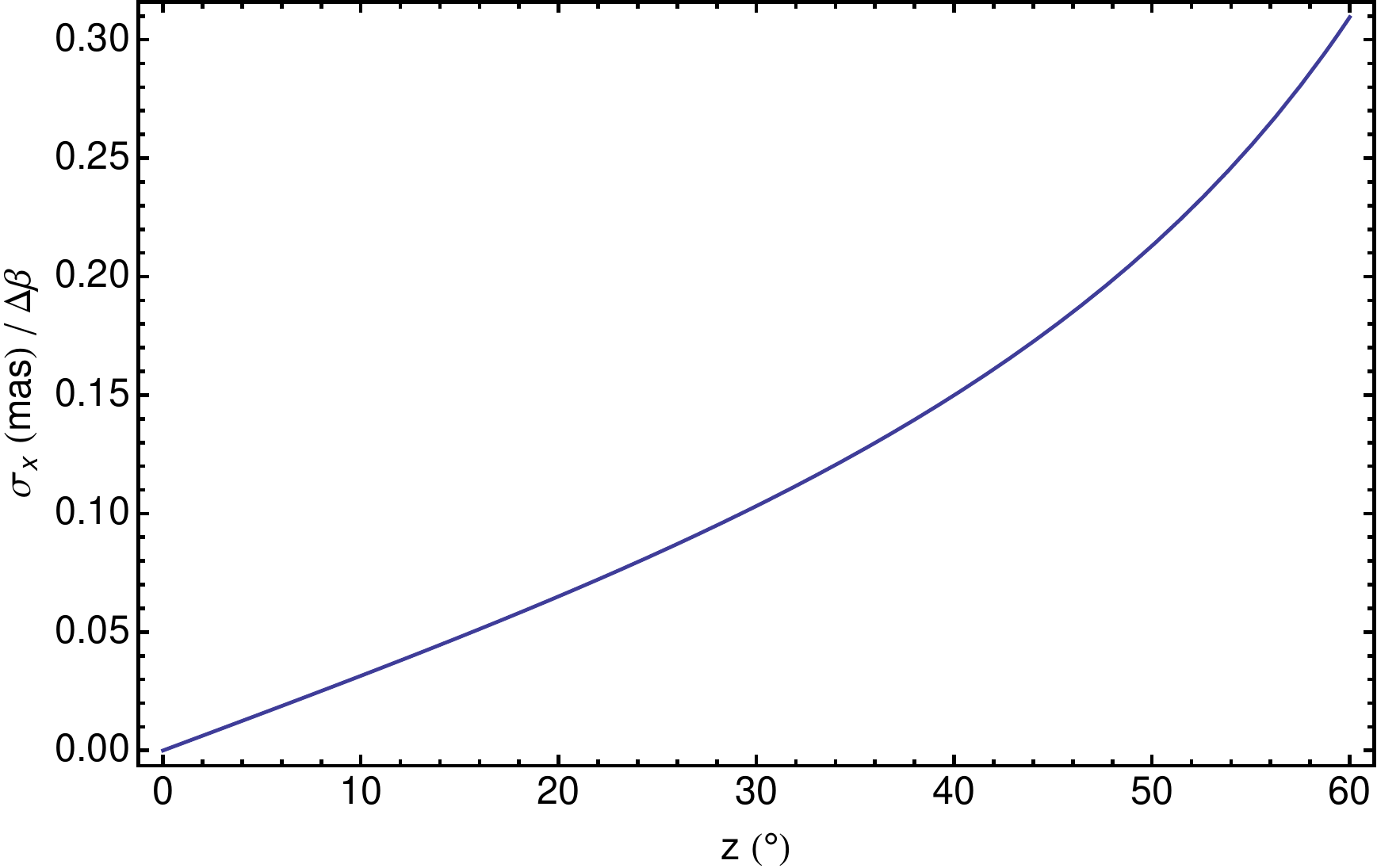}
  \caption{Positional shift in K$_S$-band of a power law type spectral energy distribution source (like Sgr~A*) per change in power law slope $\Delta \beta$, $\nu S_\nu \sim \nu^\beta$, as a function of zenith angle as induced by the chromatic nature of refraction in Earth's atmosphere.}
\label{f7}
\end{figure}

Finally, it is worth investigating the chromatic effects for Sgr~A*, the MBH itself, which in the NIR is a variable source powered by the synchrotron emission of relativistic electrons and possibly changes its power law index $\beta$ ($\nu S_\nu \sim ~\nu^\beta$) with flux \citep{Eisenhauer:2005p117, Gillessen:2006p160, Krabbe:2006p1515}, but see also \cite{Hornstein:2007p1513}.
The astrometric effect $\sigma_x$ per unit change of $\beta$ is shown in figure~\ref{f7} as a function of zenith angle. The effect reaches $\approx 200\,\mu$as for a zenith angle of $\gtrsim30^\circ$ and a change of $\Delta \beta = 2$. This is still a bit smaller than the currently achieved astrometric accuracy, but it can be a big systematic effect for future studies using extremely large telescopes. The easiest way to overcome the problem probably is the use of a narrow band filter,
but also atmospheric dispersion compensators are feasible.   
Interferometric studies of Sgr~A* \citep{Eisenhauer:2008p2000} will use several spectral channels over the K-band. In each channel, the effective wavelength will be altered by a change in color of Sgr~A*. Using five spectral pixels of $\Delta \lambda = 0.1\,\mu$m will already lower the astrometric effect to $<10\,\mu$as, and in addition a first order correction will be possible because the color of the source can be measured simultaneously.

\subsubsection{Influence of Strehl ratio}
\begin{figure}
\includegraphics[width=80mm]{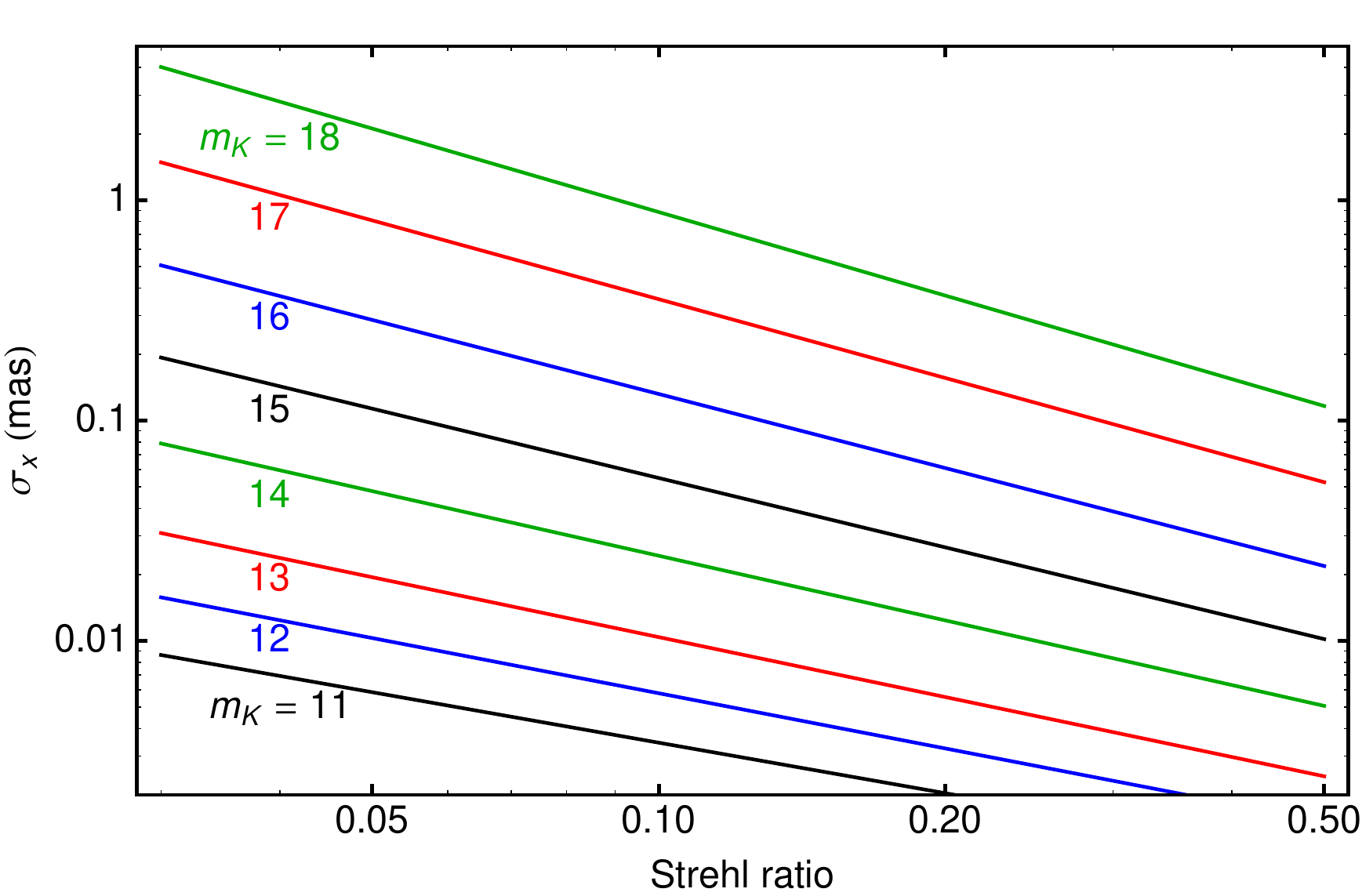}
  \caption{SNR induced position error in a K$_S$-band mosaic of 96 single frames as a function of Strehl ratio for stars of magnitudes between 11 and 18. The darker a star is, the more its SNR induced position error is sensitive to the atmospheric conditions.}
\label{fstrehl}
\end{figure}

For any given AO image, it is common practice to characterize its quality by the SR. It is strongly correlated with the FWHM of the PSF, and hence from equation~(\ref{e1}) follows that the SR limits the reachable accuracy. We simulated the influence of the SR by placing measured PSFs into noisy background maps and finding these isolated stars back using starfinder. The resulting errors as a function of SR followed very well power laws with power law indices between $-0.7$ and $-1.3$ for stars between $m_\mathrm{K}=11$ and $m_\mathrm{K}=18$ (figure~\ref{fstrehl}). The power laws get steeper with increasing magnitude since the readout noise is more important for faint sources. As a consequence, the SNR induced position error for fainter stars is more sensitive to the observing conditions than for brighter stars. 
As suggested by figure~\ref{fstrehl} and in practice, the SR induced position errors are smaller than other error terms, most notably the halo noise (section~\ref{halon}). Hence, our current data sets are not SNR limited.

\subsection{Instrumental Limitations}
\subsubsection{Distortion}
\label{r1}
Each imaging system suffers to some degree from optical distortions. These can either be a global effect for the image (such as classical optical aberrations) or small scale imperfections. The global terms should mainly be absorbed by the third order transformation between pixel positions and astrometric positions, but a floor of residuals is unavoidable. If we would use only one pointing position, these residuals would not influence the precision of the data, but would of course affect the accuracy. However, we use various pointing settings from run to run and even during one run (to increase area coverage). The residual image distortions effectively act like a random error.  

Our $13\,$mas/pix observations are typically dithered in a scheme that four positions, offset by 3.5'' from Sgr~A* in both axes, are obtained. Since the field of view is $\approx 13.5''$, the central $\approx 7''$ are covered by each frame. From these data, we usually construct multi-pointing mosaics. The effect of the image distortions here is a) that each source gets broadened slightly and b) that a superposition of the global distortions is present in the area which is covered by several pointings. Both the area of interest and the reference stars are taken from the central 7'' which are covered by all pointings, and hence the resulting distortion pattern might be a complicated function, which we represent by a cubic transformation with 20 parameters.

\begin{figure}
\includegraphics[width=80mm]{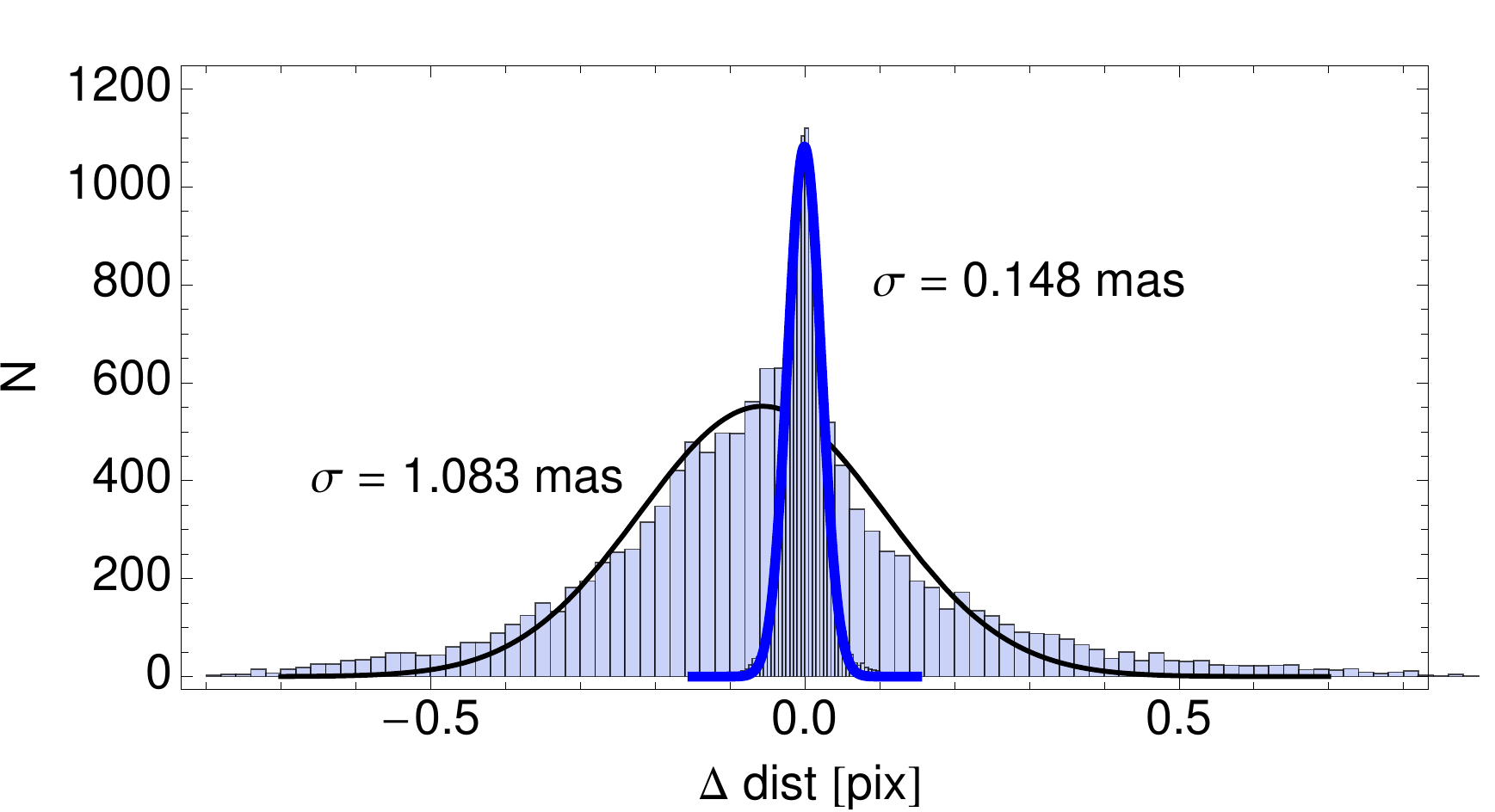}
  \caption{Compensation of image distortions by means of a cubic transformation for the data from 13 March 2008. The broad histogram is the distribution of differences of bona fide stellar positions from four pointing positions with a dither offset of 7'' after having mapped the positions onto each other with a full linear transformation. If the stars are mapped onto each other by a cubic transformation the same differences yield a much narrower distribution since image distortions are removed. The corresponding one-dimensional position errors $\sigma_x$ are given, too.}
\label{fdistcorr}
\end{figure}

Following the approach in \cite{Gillessen:2009p1117} we estimate the size of the effect for the data from 13 March 2008 by creating four single-pointing mosaics and comparing the distances of stars common in all pointings after having transformed them with a cubic transformation onto each other (figure~\ref{fdistcorr}).
The width of distribution of position differences then measures the residual image distortions, but includes also the effects of PSF uncertainty and differential tilt jitter. We obtained a value of $\sigma_x=148\,\mu$as, of which $\approx 50\,\mu$as are due to the uncertainty in the PSF, 
 $\approx 40\,\mu$as due to anisoplanatism, and $\approx 125/\sqrt{24}\,\mu$as$\,\approx 25\,\mu$as due to differential tilt jitter. The SNR induced position error is $<5\,\mu$as here, since the mean magnitude for the set of stars used was $m_\mathrm{K}=13.1$.
Hence, we conclude that the residual image distortions affect the astrometry at a level of $\approx 130\,\mu$as. This value is smaller than the corresponding number in \cite{Gillessen:2009p1117} who had not taken into account the other effects subtracted here and who were looking at a different data set from a time before the instrument got realigned.

\begin{figure}
\includegraphics[width=80mm]{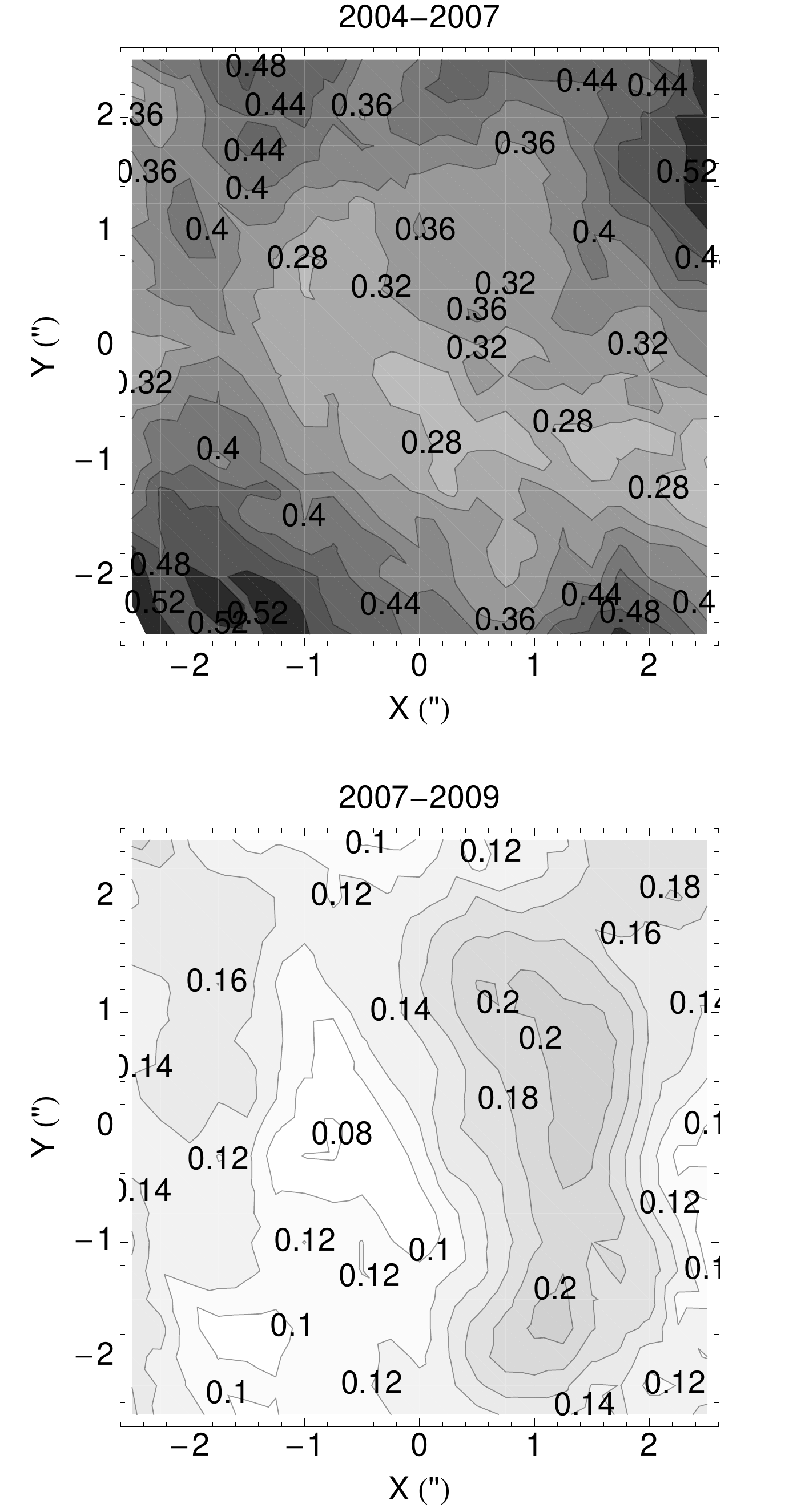}
  \caption{Maps of estimated residual image distortions. Contour labels are in mas. Top: For the period between June 2004 and fall 2007. Bottom: For the period after the realignment in 2007 until 2009. Note that identical color maps have been used, pronouncing the fact that the residual distortions are much less severe after the realignment.}
\label{fdistmap}
\end{figure}

A different route to estimate the residual image distortions is to compare the high order terms of the transformations between several epochs. For this purpose we calculated the inverse of the
third order transformations for each of the 73 NACO epochs with $13\,$mas/pix data to convert a regular grid on sky back to the respective pixel positions in each frame. Then we mapped these pixel positions to a common epoch with a full linear transformation which was determined from the reference stars for the given epoch. Like this we take into account image shifts, rotations, scale changes and shear terms and therefore the resulting jitter of the grid points measures the variations of the higher order distortions. Since the set of reference stars on average still might be moving in a linear fashion we actually determined the jitter as the residuals to a temporal linear fit for each grid point. We broke up the entire data set into three periods, corresponding to three hardware configurations: From 2002 to June 2004 a different detector was used than in the following years, and NACO underwent a major realignment late 2007. From the residuals we constructed maps of the residual distortions (figure~\ref{fdistmap}). For the first two periods, a typical value for the central arcsecond is $\sigma_x \approx 300\pm50\,\mu$as, while for the last period a significant improvement to $\sigma_x \approx 120\pm30\,\mu$as is observed. This estimate agrees with the number quoted in the previous paragraph for the data set from 13 March 2008.

For the $27\,$mas/pix camera of NACO, the residual distortions in absolute terms are higher. The procedure of \cite{Trippe:2008p1123} leads to residuals of order $1.2\,$mas, (see their figure~2, bottom; or figure~6 in \cite{Gillessen:2009p1117}). Over time, the distortion appears stable, given that the distortion parameters of \cite{Trippe:2008p1123} are consistent with each other for all data analyzed since 2002. For a distance of 10'' from the optical axis (which differs from the center of the detector), the uncertainty of the geometric model is $0.4\,$mas and hence is smaller than the residuals. We have not analyzed further, what limits the positional accuracy for the $27\,$mas/pix data. For our data sets, the residuals of this plate scale matter when the positions need to be related to an absolute coordinate system, for example for answering the question where radio-Sgr~A* is located on the frames. For many questions, however, this does not matter, e.g. when one is trying to detect accelerations. For a more detailed discussion see \cite{Gillessen:2009p1117}.

\subsubsection{Detector non-linearity}
Nominally, the current NACO detector is regarded as a linear device up to 2/3 of its capacity, corresponding to 14000 ADU for the readout mode of our data sets. Using a set of flat exposures with varying exposure times, we checked the linearity behavior of the detector and found that at 10000 ADU the non-linearity is $7\%$, at 7000 ADU $4\%$ and at 5000 ADU still $2\%$. This would in turn mean that a PSF estimate is worse for brighter stars, possibly leading to positional biases. Hence, we calculated a pixel-wise map of polynomial coefficients to correct the effect, which is now applied to all data. That correction
doubles the usable dynamic range of the detector and thus allows the use of brighter stars. It is now linear to $2\%$ up to
12 000 ADU, which for the data from 13 March 2008 corresponds to $m_\mathrm{K}=10.7$.

\begin{figure}
\includegraphics[width=80mm]{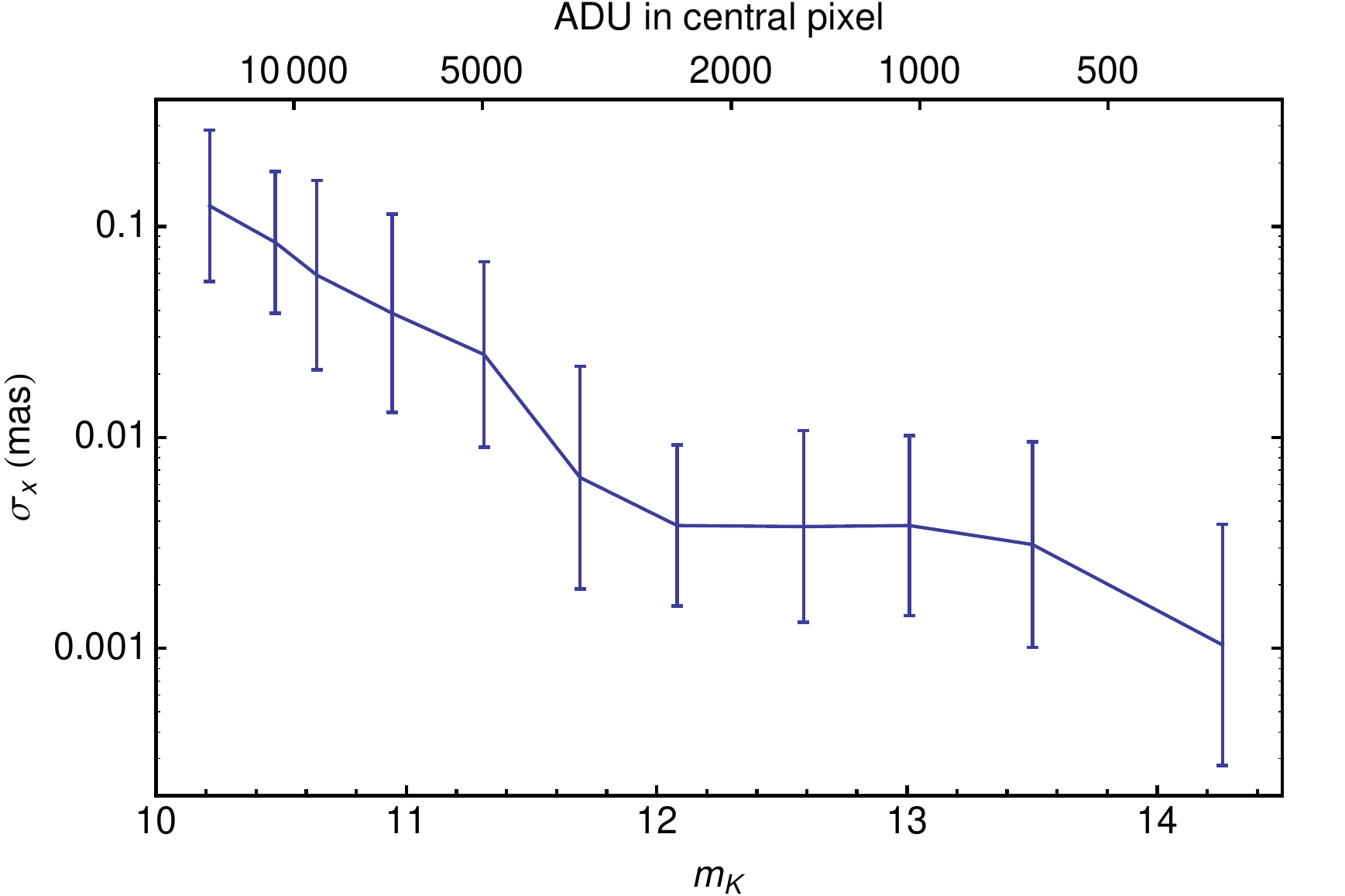}
  \caption{Position error induced by the detector non-linearity under good observing conditions as a function of flux in ADU (top axis) and stellar magnitude (bottom axis) assuming an image with DIT$=17.2\,$s. The error bars show the standard deviation of the position errors and hence measure the scatter per bin.}
\label{f10}
\end{figure}

We also tested for the potential astrometric bias by comparing the positions from a linearity-corrected and an uncorrected frame. Figure~\ref{f10} shows that the effect in a mosaic for the brightest S-stars with $m_\mathrm{K}\approx 14$ is $\lesssim50\,\mu$as and decreases for fainter stars. 

Stars that are affected by saturation are offset sporadically up to 100$\times$ the usual positional error, i.e. in the multi-mas range. Such a big effect is easily detected in the data and hence such cases can be flagged as outliers. We thus don't assign a formal error here but rather expect that for some data sets, for bright stars saturation effects make the positions unusable. Furthermore, heavily saturated stars (showing for example a flux plateau at the highest flux value) are not described at all by the PSF estimate; such stars are even recognized earlier, since the deconvolution does not yield a point-like structure in these cases.

\subsection{Analysis Limitations}
\subsubsection{Knowlegde of the PSF}
\label{secPSF}
Astrometry in a crowded stellar field necessarily needs as input an estimate for the PSF. Just using local maxima in the image as position estimates is not sufficient, since essentially all stars are located on the seeing halo of neighboring sources. This yields a background flux with a gradient and thus biases the positions of local maxima. Even worse, the background will vary with the AO correction from image to image and lead to a position jitter. Moreover, very faint stars in a seeing halo of a bright star do not produce a local maximum (but only change the shape of the seeing halo) and would be missed by an algorithm looking for local maxima.

These problems are overcome by using the knowledge of the PSF, which can be obtained from the image itself with an iterative procedure, e.g. using starfinder \citep{Diolaiti:2000p1984}. Given a certain PSF estimate, there are two basic options to obtain position estimates.
\begin{itemize}
\item {\bf PSF fitting.} A program like starfinder \citep{Diolaiti:2000p1984} iteratively fits local maxima to the image until the image is sufficiently well represented by the set of sources found, each of which is characterized by a position and flux.
\item {\bf Deconvolution.} The image on the detector is the convolution of the PSF with the original object distribution. Since a convolution in image space corresponds to a multiplication in Fourier space, the inverse operation deconvolution in the Fourier domain is a division. The presence of noise and the limited range of spatial frequencies complicate the actual algorithm. Various methods are available and have been compared \citep{Ott:1999p174}. We mostly use the Lucy-Richardson algorithm \citep{Lucy:1974p216} for astrometry.
\end{itemize}
Since both methods use the same input, one does not expect a major difference in astrometric performance. The fact that in figure~\ref{f4} the starfinder errors are larger than the corresponding deconvolution errors might reflect a non-optimum user choice for the parameters of the starfinder algorithm and/or it might be due to the differences how the non-stellar background is dealt with. 

The fact that the input PSF is not perfectly known but has to be estimated needs to be accounted for in the error budget. In order to quantify the error, we used various sets of PSF stars and compared the resulting positions. Also, we compared the positions depending on the number of iterations used for the extraction of the PSF. From the scatter of the positions (after having transformed them linearly onto each other) we conclude that the uncertainty of the PSF knowlegde yields a positional error of $50\,\mu$as.

\subsubsection{Transformations}
The fact that we need to transform the measured pixel positions with a set of reference stars is an error source because the transformation parameters also have errors. Fortunately, the corresponding statistical error can be lowered to a very low level. If $N$ stars are used as reference stars, each with a typical position error of $\sigma_x$, the additional error for the target sources due to the transformation is of order $\sigma_x/\sqrt{N}$. By choosing a large enough sample of reference stars, this statistical error can thus be lowered to any desired value as long as a sufficient number of reference stars is available. Of course, this holds only if the reference stars can all be chosen from within the field in which the PSF can be treated as constant.
Recently \cite{Gillessen:2009p1117} used $N\approx 100$ reference stars in the GC field, which suppresses effectively the transformation error to below $50\,\mu$as. Fomer studies \citep{Schodel:2002p153}
had used $N\approx 10$ reference stars, which hence at the time was contributing at the few 10\% level to the errors.

\subsubsection{Deconvolution}
\label{deconoise}
We also investigated systematically the number of iterations to be used with our implementation of the Lucy-Richardson algorithm. Using a set of 66 well isolated stars in the data set from 20 July 2007 we obtained pixel positions for different deconvolution depths. 

For less than 500 iteration steps, the resulting positions are biased up to $0.5\,$mas 
for bright stars compared to the positions in the undeconvolved image and compared to more deeply deconvolved images.  For different stars the bias acts in different directions, and hence globally it appears to be a scatter of up to $0.5\,$mas. For 2000 steps, both bias and scatter for bright stars drop below $120\,\mu$as. The optimum is reached for $\approx 10000$ iterations; also fainter sources are unbiased with that number of iterations. Much larger values (like 50000 steps) 
tend to put the flux of any star into a single pixel, which of course corresponds to larger positional errors again. The optimum depth in practice is between 5000 and 15000 steps, with better data needing less deep deconvolution.

\subsubsection{Halo noise}
\label{halon}
Visual inspection of the deconvolved frames shows that the background in the images is not flat but rather has a noise-like structure, see for example figure 1 in \cite{Gillessen:2009p1117}. The deconvolution algorithm does not discriminate between sources and background and therefore any extended light present gets concentrated into this floor of faint local maxima. 
That light is due to the imperfections in the knowledge of the PSF wings (the dominant source) and real, physical extended background light as expected from interstellar gas. Astrometrically, the spurious deconvolution peaks act as a noise source. It turns out that this halo noise is the main limitation for fainter sources. Experimentally indistinguishable from this is the confusion noise due to unresolved background sources (section~\ref{conf}). 

While the term halo noise describes the influence of the imperfectly known PSF halo of surrounding stars, the PSF uncertainty discussed in section~\ref{secPSF} refers to the effects of the imperfectly known PSF on the source itself. The PSF uncertainty from section~\ref{secPSF} would be present also fur sufficiently isolated sources, the importance of the halo noise is a consequence of the stellar crowding. It is also worth noting that the halo noise is not specific to the deconvolution technique. Also a starfinder-like algorithm is not free from that noise source. It generally arises from the fact that the seeing halo of neighboring, brighter stars cannot be subtracted perfectly, which owes to the large dynamic range in the images and the high surface density of stars.

As a first test to estimate the effect of halo noise, we added Gaussian peaks with the same width as the typical width for real stars into the deconvolved frame from 13 March 2008 at random positions, avoiding positions at which the added source would overlap with an already existing and detected star. We then re-identified the random sources and determined their positional uncertainty by calculating the median deviation of the distribution of differences between the respective initial and re-identified positions. This tests by how much simulated stellar sources are shifted due to the combined deconvolution and confusion noise. The resulting positional error can be described very well by a relation of type
\begin{equation}
\sigma_x =  \Sigma_x(r) \,\times 10^{0.4(m_\mathrm{K}-14)} \,\,,
\label{esigx}
\end{equation}
where $\Sigma_x(r)$ is a characteristic error for a source with $m_\mathrm{K}=14$ at a given radial distance $r$ from Sgr~A*. In the chosen data set $\Sigma_x$ has the following radial dependence: 
\begin{equation}
\begin{array}{l|cccccccc}
r['']&0.22&0.47&0.77&1.07&1.36&2.07&3.18&M\\
\hline
\Sigma_x[\mu\mathrm{as}]&273&176&122&133&110&105&82&63
\end{array}\nonumber
\label{esigx2}
\end{equation}
The last bin, labelled $M$, corresponds to manually selected, well-behaved background regions in the deconvolved image. There are two reasons why the deconvolution and confusion noise decreases with radius: a) The unrecognized, faint stellar population falls off with radius. b) The stray light, leading to a variable background on top of the PSF, follows the general trend that the light is roughly concentrated towards Sgr~A*. 

\begin{figure}
\includegraphics[width=80mm]{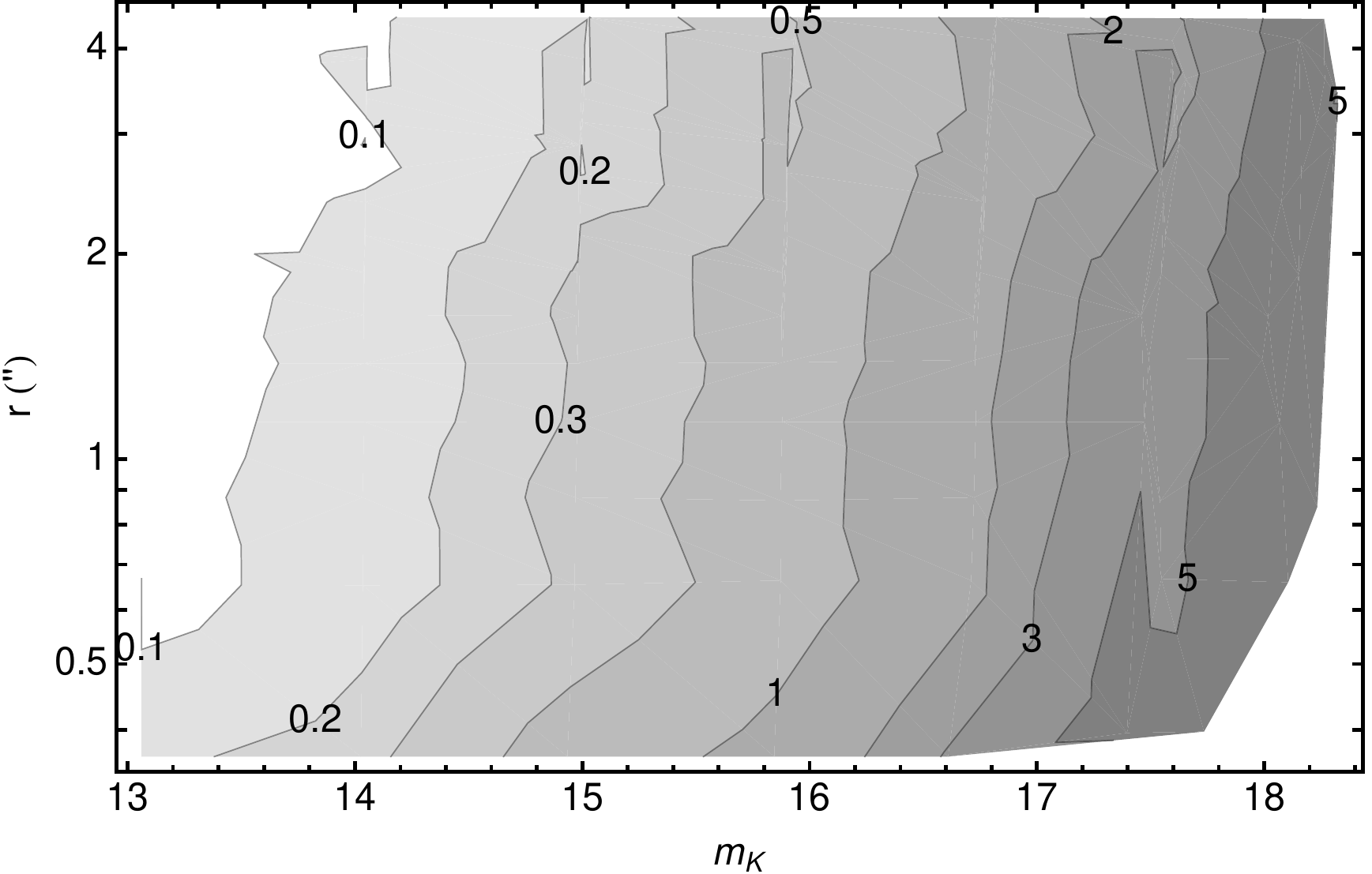}
  \caption{Position errors due to halo (and confusion) noise as a function of stellar magnitude and distance from Sgr~A*. The errors are obtained by adding simulated stars to an undeconvolved frame and finding the sources back after deconvolution. The contour label units are mas.}
\label{f13}
\end{figure}

In a second test we added the full PSF into the undeconvolved frame, neglecting the effects of anisoplanatism. The modified frame was then deconvolved as before and stars were identified, yielding again an estimate for the error introduced by the deconvolution with imperfectly known background and PSF halo. This procedure yielded the errors as a function of magnitude and radius from Sgr~A* as shown in figure~\ref{f13}. The numbers obtained are broadly consistent with the results from the first test where Gaussian peaks were added to the deconvolved frames.

Hence, the combination of halo and confusion noise is a severe limitation. Only for the brightest S-stars ($m_\mathrm{K}\approx 14$) it gets as small as the errors due to residual image distortions. In the next section, we show that the confusion noise actually contributes much less than the halo noise and we have to conclude that halo noise is limiting the astrometric accuracy for stars fainter than $m_\mathrm{K}\approx 15$. 

Due to the importance of the halo noise, we have repeated the last test (putting full PSFs into an existing image and finding the stars back) using PSF fitting (starfinder) instead of Lucy-deconvolution. We used 100 artificial stars per magnitude bin on the image from 13 March 2008, the positional error again is the width of the distribution of the differences between detected and simulated position. The results are very similar to what we found with deconvolution. The positional uncertainty using starfinder
is slightly larger than for the deconvolution, similar to the findings in figure~\ref{f4}. The photometric correctness on the other hand is slightly better for the PSF fitting method. This can be understood, since the Lucy algorithm tends to pull flux fro the surroundings into local maxima during the iterative procedure. For a low SNR source this can bias the flux estimate. Overall, we conclude that halo noise is an important noise source, and it is not specific to the method used to extract positions.

\subsection{Astrophysical Limitations}
\subsubsection{Confusion}
\label{conf}
Astrometry in the GC is affected by the stellar crowding close to the MBH. The S-stars reside in a cusp with a steep density profile. As a consequence, the closer a star is to the MBH, the more frequently it is confused with some other source. Not all confusion events are recognized, and therefore unrecognized confusion events will contribute to the positional error budget.
This error is already included in the discussion of section~\ref{halon}. It is nevertheless instructive to single out the effect of the stars, since that limit would remain even if the instrumental halo noise could be removed.

As an extreme example for a possible confusion error, it is worth mentioning that the orbit of the star S2 possibly was affected in 2002, during its pericenter passage, by such an event. Both recent analyses \citep{Ghez:2008p945,Gillessen:2009p1117} therefore treat the respective 2002 data separately; either by ignoring it or by assigning large errors to it.

\begin{figure}
\includegraphics[width=80mm]{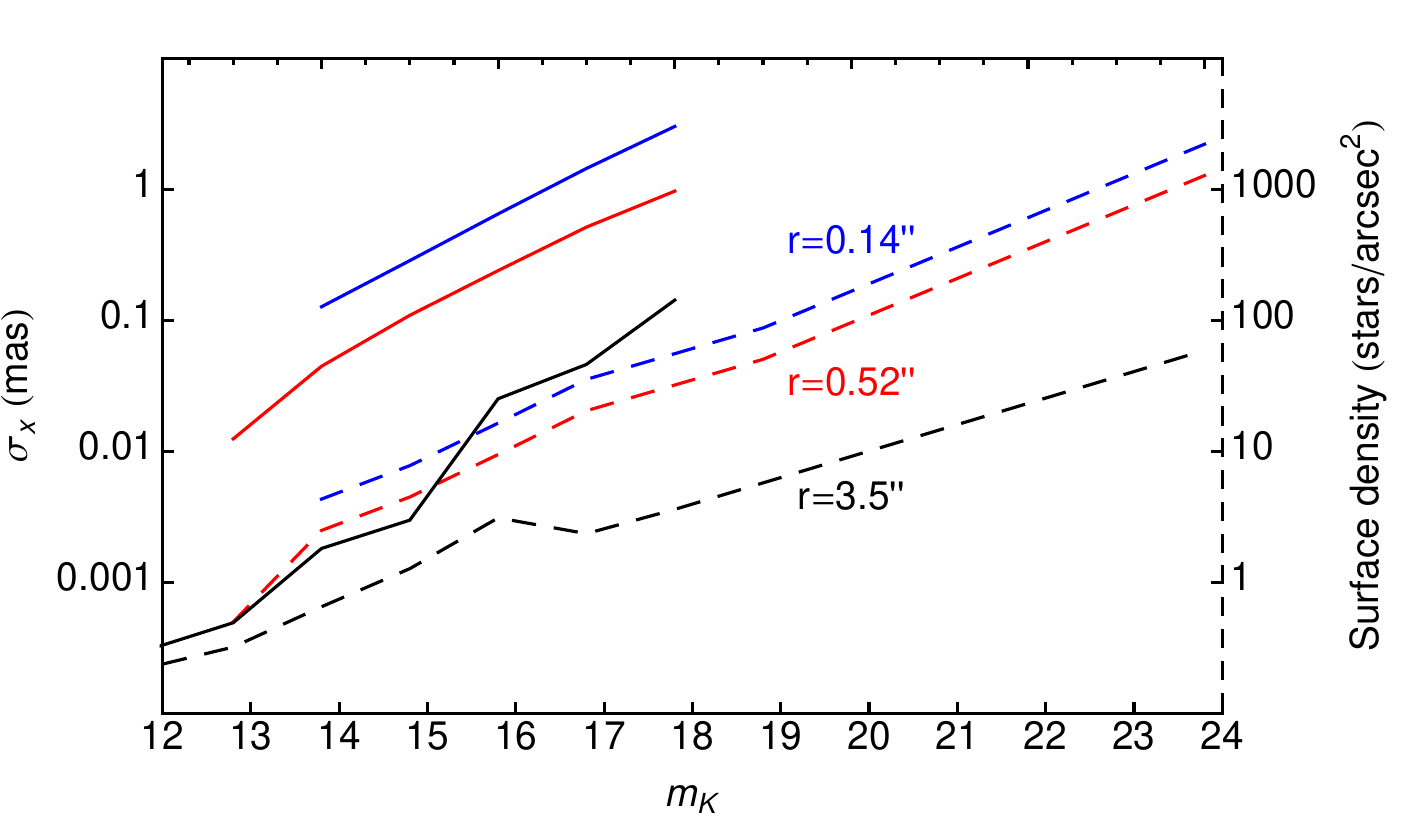}
  \caption{Assumed stellar densities and resulting position errors due to confusion noise: Dashed lines and right axis: assumed stellar surface density as a function of magnitude. Solid lines and left axis: resulting positional errors due to unrecognized confusion events with fainter stars. The blue lines are for $0''<r<0.2''$, the red for $0.2''<r<0.8''$ and the black for $r=3.5''$. }
\label{f8}
\end{figure}

In order to assess the magnitude of the confusion noise, we simulated stellar background populations in a Monte-Carlo fashion. That needed two basic input distributions: a K-band luminosity function and the radial surface density profile. We based these on the findings of \cite{Genzel:2003p151}. We used three radial bins for our simulations: $0''<r_1<0.2''$, $0.2''<r_2<0.8''$ and $r_3=3.5''$. Since we had to assume the density also for stars much fainter than what can be measured, some assumptions had to be used. For $r_3$ we extrapolated the cluster KLF of  \cite{Genzel:2003p151} down to $m_\mathrm{K}=24$. For $r_{1,2}$ and $m_\mathrm{K} \le 18$ we used the KLF as estimated from the S-stars cusp, scaled to the respective expected surface density from
the radial profile. At fainter magnitudes we extrapolated with a KLF from the same radial region that only counts stars which are not identified as late-type stars. This essentially assumes that for $m_\mathrm{K}>18$ only main sequence stars are present. Figure~\ref{f8} shows the densities used.

Using the assumed densities per magnitude we simulated stellar fields, using Gaussian profiles with a FWHM of $30\,$mas\,$-\,42\,$mas (magnitude dependent, to mimic the FWHM of real sources in the deconvolved frames) and a binning of $13\,$mas/pix. For each image, the target star was placed in the center of a box of 12 pixels width and background stars fainter than the target star were added. The positional error of the target star is given by the difference between input position and the position at which it is found back by fitting the star with a Gaussian profile plus a floor. We used up to $10^5$ simulated images per magnitude bin. The distribution of positional differences per magnitude bin was then fit with a Gaussian, the width of which estimates the position error. The resulting errors as a function of magnitude are shown in figure~\ref{f8}.

For the S-stars cluster ($r\lesssim0.8''$) the error due to unrecognized confusion is of similar magnitude as the error due to halo noise. For larger radii, the confusion induced error is smaller than the halo noise as the stellar densities drop rapidly with radius. The increase of confusion error with stellar magnitude is well described by a power law of type $C \times 10^{0.4\, m_\mathrm{K}}$ (as the halo noise).

\subsubsection{Relativistic Effects}
Currently, relativistic effects have not yet been detected in the data of any star orbiting the GC MBH. Actually, detecting $\beta^2$ effects will be possible probably first in radial velocity measurements \citep{Zucker:2006p194}. Astrometrically detectable deviations from Newton's law have not yet been seen in the GC. Vice versa, the effects can currently be neglected in the analysis. Still, a few effects are worth discussing here.
\begin{itemize}
\item {\bf Gravitational light deflection by the MBH}. The deflection angle for a star sufficiently far behind
the MBH (at distance $z$) can be approximated by
\begin{equation}
\theta\approx 20\,\mu\mathrm{as}\,\frac{z}{b}\,\,
\label{e6}
\end{equation}
where $b$ is the impact parameter \citep{Nusser:2004p1979}. For smaller values of $z$, $\theta$ decreases compared to equation~\ref{e6}, and for $z=0$ it is half of it. Generally the effect leads to very faint secondary images ($m_\mathrm{K}\approx22$ at best), and lensing only can become important when a star crosses the line of sight to Sgr~A* behind Sgr~A*. Currently none of the stars tracked is in that regime and we neglect the effect. On the other hand, detecting the effects of secondary images would be extremely interesting. \cite{Bozza:2005p129,Bozza:2009p1970} calculate the appearance of the secondary images for the known orbits. It will be extremely difficult to detect these faint sources in the crowded field around Sgr~A* ever. Another application would be the idea of \cite{Alexander:2001p1204} who suggested to use the secondary images of background stars to pinpoint the position of the MBH.
\item {\bf Gravitational light deflection by the Sun}. 
Light deflection by the gravitational field of the sun can be approximated by \citep{Lindegren:2006p1980}
\begin{equation}
\theta\approx 4\,\mathrm{mas}\times\,\cot \Psi/2\,\,,
\label{e7}
\end{equation}
where $\Psi$ is the angle between Sun and the observed object. While the absolute magnitude of the effect is as large as $22\,$mas for $\Psi=20^\circ$ (the minimum useful distance), the differential effect (relevant for relative astrometry as in our case) over a field of view of 20'' is always well below $10\,\mu$as and would even mainly be absorbed in the linear terms of the transformations. 
\item {\bf Differential aberration}. The classical light aberration yields for a small field of view effectively
a change in plate scale in one direction \citep{Lindegren:2006p1980}. The effect will be absorbed into the parameters of the transformation when converting pixel positions to astrometric positions via reference stars if and only if one allows also for the linear shear terms in the transformation. Otherwise, a positional error of
\begin{equation}
\theta\approx \Theta_\mathrm{FoI}\,\frac{v}{c}\,\cos\Psi\,\,,
\label{e8}
\end{equation}
can occur, where $v\approx30\,$km/s is the speed of Earth and $\Psi$ the angle between the apex point and the target. Numerically, this evaluates to $1\,$mas$\,\times\cos\Psi$ for a field of interest of $\pm2''$ and hence would be a huge effect.
It is, however, absorbed into the shear terms of the transformation.
\end{itemize} 
\section{Summary \& Conclusion}
\begin{figure}
\includegraphics[width=80mm]{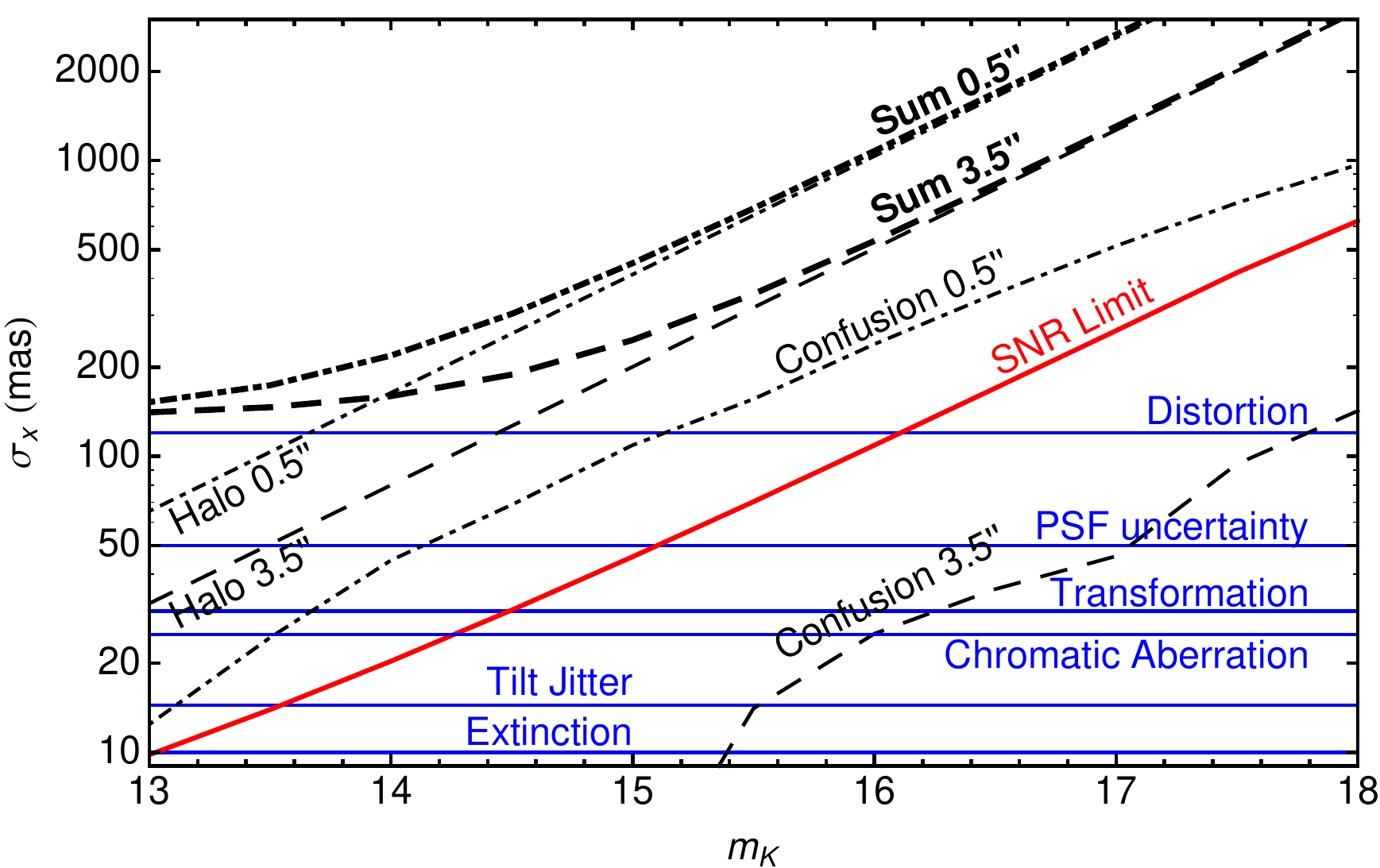}
  \caption{Summary of error sources as a function of stellar magnitude. The blue, horizontal lines
  are constant contributions. From top to bottom these are: residual image distortions, PSF uncertainty, coordinate transformations, differential chromatic aberration, differentail tilt jitter and extinction uncertainty. The red, diagonal, solid line is the SNR limit. The dot-dashed curves are for the central cusp ($r=0.5''$) and show from bottom to top the confusion error, the halo error and in thick the sum of the errors for that radius. The dashed curved are the same for $r=3.5''$.}
\label{f9}
\end{figure}

We have analyzed a multitude of error sources that potentially influence and bias stellar positions as obtained from adaptive optics assisted imaging data in crowded stellar fields. The domain of application in this work is the stellar cusp in the Galactic Center. Figure~\ref{f9} summarizes the various error sources for a mosaic of 48 frames with DIT$=17.2\,$s and NDIT=2 using NACO in K-band and the $13\,$mas/pix image scale, assuming the current instrumental setup and good observing conditions.

We find that for stars fainter than $m_\mathrm{K} \approx 15$ the main error source for position measurements is halo noise, resulting from the imperfectly subtracted PSF seeing halos of surrounding, brighter stars. For stars of $m_\mathrm{K} \approx 14$ and brighter, the main limitation is residual image distortions and the uncertainty in the PSF. In order of importance for a star with $m_\mathrm{K} = 15$ in the central arcsecond, we estimate the following error contributions:
\begin{itemize}
\item Halo noise: $300\,\mu$as
\item Residual image distortions: $130\,\mu$as
\item Confusion noise: $100\,\mu$as
\item PSF uncertainty: $50\,\mu$as
\item SNR induced position uncertainty: $45\,\mu$as
\item Coordinate transformations: $30\,\mu$as
\item Differentail tilt jitter: $15\,\mu$as
\item Extinction uncertainty: $10\,\mu$as
\item Differential chromatic aberration: $10\,\mu$as
\item Detector non-linearity: $<5\,\mu$as
\end{itemize}

It is worth noting that our figure~\ref{f9} cannot be compared to figure 3 of \cite{Ghez:2008p945}, who only show the centroiding uncertainties and not the astrometric accuracy. The proper comparison is with our figure~\ref{f4}.

Given the dominance of the halo noise in the error budget, the most promising route to improve the astrometry on existing data is to get better PSF estimates, in particular for the large seeing halo. This task is complicated due to the stellar crowding which is so dense that essentially no star has an unperturbed halo. It may also be worthwhile to think about blind deconvolution, which extracts the PSF while finding stars in an iterative way and in principle could use as much information as possible from any image. The advantage would be that basically all stars are used to estimate the PSF halo, and not just the ones which a user has selected a priori. Another way to improve the PSF estimates might be the techniques of PSF reconstruction from the wavefront sensor data \citep{Gendron:2006p2259}.

For the brightest stars ($m_\mathrm{K}\approx14$), the next step of improvement would come from a better handling of residual image distortions. One possible route could be to actually map the distortions and then apply a correction to the pixel positions obtained from the data. For that, it is also important to have a well designed, well aligned and stable optical system. 

Other, desirable improvements which could be implemented in the analysis using current data are: a) take into account anisoplanatism during the deconvolution process (moderately important), b) correct for the atmospheric chromatic bias due to the different stellar types (important in H-band only) and c) extend the linearity correction to earlier data obtained with another detector (barely important).

It should also be noted that most of the errors effectively act as random errors, if many epochs are considered. Hence, we expect that observing at many epochs will average out many errors discussed. Nevertheless some care needs to be taken that this actually happens. The halo noise might get correlated between epochs if consistently the same PSF stars and same PSF size get used. Using different bands, experiencing variable atmospheric conditions, the motion of stars and varying the way to estimate the PSF during the analysis help to avoid that the error turns into a bias. For the analysis, this means that from a given data set frames should be added up until the faintest stars of interest are well detected, but not beyond that. If more frames are available, it is better to create a second mosaic, thus lowering the halo noise induced error. The only source of error which unavoidably leads to correlated position errors is unrecognized confusion events, since for typical proper motions, stars are confused for a few years. 

Our results also indicate that further instrumental improvements mainly would come from better angular resolution and higher Strehl ratios. It is somewhat trivially clear that this would yield better astrometry. The value of our analysis is mainly to show that this indeed is limiting the astrometry in the Galactic Center. Current observations are not SNR limited, for example. Also, we cannot exclude that at the level of $\approx 200\,\mu$as the astrometry has reached a limit, given current telescopes and adaptive optics systems.

Since the resolution is essentially set by the apertures of the telescopes, only extremely large telescopes (with $30\,$m to $40\,$m aperture) or interferometric means in the future will
further improve the angular resolution.

The Strehl ratio deserves a bit more discussion. Clearly, the astrometry in the Galactic Center would benefit from an adaptive optics system that yields the highest Strehl ratio possible over a relatively small field. This means that adaptive optics systems that yield a more moderate correction over wider fields are not well-suited for studying stellar dynamics close to the massive black hole in our Milky Way. Interesting for the GC seem the concepts for extreme adaptive optics systems that should deliver Strehl ratios up to 90\% over a limited region \citep{Macintosh:2003p2219,Conan:2004p2220} Finally, good astrometry will continue to depend on suitable atmospheric conditions.

In a nutshell: Further improving the astonishing measurements of stellar orbits in the GC requires larger telescopes, higher Strehl ratios, better knowledge or reconstruction of the seeing halo of the PSF, correction of residual image distortions - and ideally even a combination of these.

\bibliographystyle{mn2e}
\bibliography{papers}

\appendix

\label{lastpage}

\end{document}